\documentclass[fleqn,usenatbib]{mnras}

\usepackage{newtxtext,newtxmath}

\usepackage{amssymb}
\usepackage{pifont}
\newcommand{\cmark}{\ding{51}}%
\newcommand{\xmark}{\ding{55}}%

\usepackage[T1]{fontenc}
\usepackage{ae,aecompl}

\usepackage{graphicx}	
\usepackage{amsmath}	
\usepackage{amssymb}	
\usepackage{longtable}
\usepackage{lipsum} 

\title[Hot ISM phase generation by SNe]{Hot phase generation by supernovae in ISM simulations: resolution, chemistry and thermal conduction}

\author[Steinwandel et al.]{
Ulrich P. Steinwandel,$^{1,2}$\thanks{E-mail: usteinw@usm.lmu.de}\thanks{E-mail: uli@mpa-garching.mpg.de}
Benjamin P. Moster,$^{1,2}$, Thorsten Naab$^{2}$, Chia-Yu Hu$^{3}$ \newauthor \ and Stefanie Walch$^{4}$
\\
$^{1}$Universit\"{a}ts-Sternwarte M\"{u}nchen, Fakult\"{a}t f\"{u}r Physik, LMU Munich, Scheinerstr. $1$, $81679$, Germany\\
$^{2}$Max Planck Institute for Astrophysics, Karl-Schwarzschild-Str. $1$, $85748$, Garching, Germany\\
$^{3}$Center for Computational Astrophysics, Flatiron Institute, $162$ $5$th Ave NY NY, USA\\
$^{4}$Physikalisches Institut der Universit\"{a}t zu K\"{o}ln, Z\"{u}lpicher Strasse $77$, $50937$ K\"{o}ln, Germany
}

\date{Accepted XXX. Received YYY; in original form ZZZ}

\pubyear{2019}

\begin{document}
\label{firstpage}
\pagerange{\pageref{firstpage}--\pageref{lastpage}}
\maketitle

\begin{abstract}
Supernovae (SN) generate hot gas in the interstellar medium (ISM), help setting the ISM structure and support the driving of outflows. It is important to resolve the hot gas generation for galaxy formation simulations at solar mass and sub-parsec resolution which realise individual supernova (SN) explosions with ambient densities varying by several orders of magnitude in a realistic multi-phase ISM. We test resolution requirements by simulating SN blast waves at three metallicities ($Z = 0.01, 0.1$ and $1 Z_{\odot}$), six densities and their respective equilibrium chemical compositions ($n=0.001$ cm$^{-3}$ - $100$ cm$^{-3}$), and four mass resolutions ($0.1$ - $100$ M$_{\odot}$), in three dimensions. We include non-equilibrium cooling and chemistry, a homogeneous interstellar radiation field, and shielding with a modern pressure-energy smoothed particle hydrodynamics (SPH) method including isotropic thermal conduction and a meshless-finite-mass (MFM) solver. We find stronger resolution requirements for chemistry and hot phase generation than for momentum generation. While at $10$ M$_{\odot}$ the radial momenta at the end of the Sedov phase start converging, the hot phase generation and chemistry require higher resolutions to represent the neutral to ionised hydrogen fraction at the end of the Sedov phase correctly. Thermal conduction typically reduces the hot phase by $0.2$ dex and has little impact on the chemical composition. In general, our $1$, and $0.1$ M$_{\odot}$ results agree well with previous numerical and analytic estimates. We conclude that for the thermal energy injection SN model presented here resolutions higher than  $10$ M$_{\odot}$ are required to model the chemistry, momentum and hot phase generation in the multi-phase ISM.
\end{abstract}




\section{Introduction}
\label{sec:Intro}

The galactic interstellar medium is shaped by the feedback of massive stars. A physical process which is of paramount importance in this picture are SN-explosions. Core-collapse-supernovae (SNe) eject gas at supersonic velocities of several thousand kilometres per second \citep[e.g.][]{Blondin1998, Janka2012} and drive blast waves into the ISM. The evolution of SN-remnants has been extensively investigated analytically, and can be divided in four phases \citep[e.g.][]{Woltjer1972}: the phase of 'free-expansion' which is terminated when the swept up mass equals the ejecta mass (roughly a few 100 years), the energy conserving Sedov-Taylor (ST) phase \citep[e.g.][]{Sedov1946,Sedov1959, Taylor1950}, the pressure driven snow plough (PDS) phase and the momentum conserving snow plough phase (MCS; \citealt{Falle1975}). In the PDS-phase the pressure of the bubble interior further generates momentum until pressure equilibrium with the shell is reached and the bubble evolves the remnant in the low pressure regime \citep{Cox1972,Gaffet1983, Cioffi1988, Cohen1998, Haid2016}. In the MCS-phase the momentum of the shell is conserved and the further evolution of the shell is driven by the generated inertia of the swept up mass \citep{Cioffi1988, Haid2016}. The energy conserving ST-phase is of particular importance as it is the main momentum generating phase with an expanding hot bubble. This momentum build up due the ST-phase drives turbulence within the ISM and regulates star formation within galaxies. It can last from a few $1000$ years in dense environments to several million years in low density environments. The duration of the ST-phase and the momentum and hot phase generation during this phase is highly dependent on the cooling properties of the ambient ISM. In this phase the SN-remnant behaves adiabatically because there are (nearly) no cooling losses that can remove energy from the system. 

Blast waves replenish the hot phase of the ISM \citep{Cox1974, Mckee1977} through shocks \citep[e.g.][]{Gotthelf2001}, generate momentum and build up turbulence \citep{Elmegreen2004, Scalo2004} in the warm neutral medium (WNM) of the ISM \citep{MacLow2004}. The hot phase generated by SNe fills a large fraction of the turbulent ISM volume \citep[e.g.][]{Mckee1977,Ferriere1998,Konyves2007} in which smaller and cooler clouds are embedded, as observed in the local ISM \citep[e.g.][]{Frisch2011}. SNe also enrich the ISM with metals and dust \citep[e.g.][]{Dwek1998, Indebetouw2014, Kobayashi2011, Matsuura2011}, different atomic species, and form molecules in their remnants \citep{Spyromilio1988, Grefenstette2014, Fransson2016, Kamenetzky2013}.
When multiple SNe occur in  a low density environment, subsequent explosion add mass to the shell and heat the interior of the bubble which can lead to the formation of 'super-bubbles' \citep{Castor1975, Weaver1977, Mccray1987, Tomisaka1986, Maclow1988, Koo1992}. Theses can drive strong galactic outflows which redistribute the gas on galactic scales and moves a fraction of it towards the circum galactic medium (CGM) of the galaxy. In more general terms, the feedback from SNe can regulate the cosmic baryon cycle and the star formation rate (SFR) of galaxies across cosmic time \citep[c.f.][]{Somerville2015,Naab2017}.

In cosmological simulations of galaxy formation, stellar feedback leads to a better agreement with the properties of the observed stellar and gaseous components, such as the low conversion efficiency of gas into stars which are determined empirically
\citep[e.g.][]{Moster2010b,Behroozi2010,Moster2013} and disk-like galaxy morphologies \citep{Naab2017, Somerville2015}. However, most simulations cannot resolve the momentum and hot medium generating phases of blast waves and rely on "sub-resolution" implementations to mimic the SN impact.
Simulations adopting these methods have been able to produce galaxy morphology and population properties in agreement with observations \citep[e.g.][]{Guedes2011, Aumer2013, Marincacci2014,Agertz2013,Wang2015,Hopkins2018}, and empirical predictions \citep[e.g.][]{Moster2018,Behroozi2018}. However, it remains unclear how well the underlying physical processes are really captured, and to what degree these results are achieved by fine-tuning model parameters. 

Recently, it has become possible to simulate low mass galactic systems like dwarf galaxies at such high resolution that the impact of supernovae from individual massive stars can be captured in the relevant temperature and density regimes of the multi-phase ISM \citep{Forbes2016, Hu2016, Hu2017,Emerick2019}. In particular studies by \citep{2019ApJ...879L..18L,2019arXiv191105093L} have extended previous galactic ISM studies to environments with much  more extreme densities and star formation rates. This approach, however, requires a detailed understanding of the resolution requirements for accurately tracking the impact of individual blast waves in the turbulent cold, warm and hot ISM. Many high-resolution simulations of SN blast waves in ambient homogeneous media or turbulent ISM patches have been carried out to understand their evolution in detail \citep[e.g.][]{Ostriker1988, Blondin1998, Thornton1998, Draine2011, Kim2015, Walch2015, Gatto2015,Martizzi2015, Badjin2016, Haid2016, Ohlin2019}. While some of these studies aim for detailed blast-wave evolution \citep[e.g.][]{Thornton1998, Badjin2016} in one or two dimensional simulations, others aim for the effects of supernova feedback in a three-dimensional approach for investigating the effects on the three-dimensional galactic ISM \citep[e.g.][]{Kim2015, Ohlin2019} or investigate the effect of multiple feedback events on the galactic ISM \citep[e.g.][]{Gatto2015}.

Apart from emerging resolution requirements, it has become clear that the distribution of ambient densities of SN blast waves play an important role for how efficiently they can drive turbulence and outflows \citep{Girichidis2016, Hu2016, Hu2017, Gatto2017, Naab2017, Seifried2018,2019arXiv191105093L}. While the turbulent component of the ISM can be generated by adopting some feedback prescription that injects the momentum of SNe, the galactic winds are driven by the pressure that is generated by the SNe within the turbulent ISM. In simulations, many factors like spatial resolution, stellar winds, radiation, SN clustering and binary or runaway stars can affect the ambient SN density distributions \citep{Kim2017, Gentry2017, Naab2017, Peters2017, Fielding2018}. However, not only numerical implementation, resolution constraints and ambient density distribution of blast waves determine their impact but also the complexity of the physical modelling. For example magnetic fields, cosmic rays \citep[e.g.][]{Diesing2018, Gupta2018} and thermal conduction \citep[e.g.][]{Keller2014,El-Badry2019} change their evolution. 

In this paper we use two particle based hydrodynamical methods: modern smoothed particle hydrodynamics (SPH) and the meshless finite mass (MFM) method \citep{Gaburov2011, Hopkins2015}. We couple both solvers to a non-equilibrium chemical network to estimate their ability to converge on blast wave evolution in cold, warm and hot ambient media at different numerical resolutions. Furthermore we use the SPH implementation to probe the impact of thermal conduction on individual blast waves. While these studies have been carried out extensively using different grid codes in one or two dimensions, results in three dimensions are rare and detailed studies for particle codes are missing. However, as the ISM is a highly structured, turbulent and multi phase fluid it is of importance to investigate the effects of supernovae on the galactic ISM at the highest resolutions we can achieve today in galaxy formation and evolution simulations in three dimensions to work out the limitations on the physics of the ISM in galactic scale simulations. We test our results against detailed blast wave studies carried out in one or two dimensions \citep[e.g.][]{Blondin1998, Thornton1998, Badjin2016} and check how well we can resolve the global evolution of the Sedov-Taylor phase with momentum injection and hot phase generation under controlled boundary conditions. We also address resolutions requirement for current and next generation galaxy formation simulations without sub-grid modelling of individual supernovae. 

The paper is structured as follows. In section \ref{sec:methods} we briefly present the numerical and physical methods that are important for this study and discuss the simulation setup and the initial parameters that are important for our cooling and chemistry network. In section \ref{sec:idealised} we discuss our results for isolated SNe blast waves in a homogeneous medium (e.g. structure of the blast wave its environmental dependence). In section \ref{sec:semi} we use the physical properties from our simulations to constrain the expectation value for a hot phase to form as a function of their environmental densities. In section \ref{sec:iso_cond} we discuss the effect of isotropic heat conduction on the SN-remnant evolution in our highest resolution runs. In section \ref{sec:galaxies} we derive the consequences of the presented supernova-feedback scheme on the ISM in galaxy scale simulations and show results for supernova-driven ISM-patches, remnants in structured media and finally full galactic disc simulations at solar mass and sub-parsec resolution. Finally, we summarise our results in section \ref{sec:conclusion}.

\section{Simulation method}
\label{sec:methods}

\subsection{Hydrodynamics}

We run our set of simulations with our version of Gadget-3 \citep{springel05}, which includes implementations for several hydrodynamics solvers, such as pressure-energy SPH \citep{Hu14} and the meshless finite mass (MFM) \citep[e.g.][]{Gaburov2011, Hopkins2015}. In \citet{Hu14} it is shown that the improved pressure-energy SPH accurately captures shocks and instabilities in several idealised test problems. While MFM shares the kernel weighted density computation with SPH, the hydrodynamical flux vectors are integrated over the one-dimensional Riemann-problem defined on the surface between two reconstruction points of the fluid equations (particles). In combination with a second order reconstruction of the flux gradients \citep{Gaburov2011, Mocz2014} it is possible to obtain a second order hydrodynamical scheme with an appropriate mathematical consistency and convergence proof as shown in \citet{Lanson2008}. We follow the slope limiting and reconstruction procedures by \citet{Hopkins2015}. We calculate the quadrature point in first order via $\mathbf{x}_{ij} = (\mathbf{x}_{i} + \mathbf{x}_{j})/2$ and solve the one dimensional Riemann-problem with a Harten-Lax-van-Leer (HLL) Riemann-solver with an approximate reconstruction of the contact wave (HLLC) following \citet{Toro1999}. The time integration scheme that we adopt follows the description of \citet{Springel2010} with a common CFL-criterion for the timestep. 
In strong shocks driven by supernova blast waves particles can move very fast and interact with other particles, which might be on much longer time steps and cannot react accurately. To avoid this we activate all nearby particles (within a kernel radius) and put them on the same short time step. The procedure is similar to the methods described in \citet{Saitoh2009} and \citet{Durier2012}.

\subsection{Thermal conduction}
\label{sec:thermal_conduction}
Some of our simulations employ a prescription for thermal conduction. We follow the implementation of \citet{Jubelgas2004} with updates for this scheme presented by \citet{Petkova2009}. We implement isotropic conduction given as a local transport process for the internal energy of a fluid tracer with the heat flux $\mathbf{j}$.
\begin{align}
    \mathbf{j} = -\kappa \nabla T,
    \label{eq:heatflux}
\end{align}
where $\kappa$ is the conduction coefficient and $T$ is the temperature. The total change of energy is
\begin{align}
    \rho \frac{du}{dt} = -\nabla \textbf{j}, 
    \label{eq:cond_energy}
\end{align}
with the thermal energy $u$ and the density of the fluid $\rho$. The conduction equation can be obtained by inserting equation \ref{eq:heatflux} into \ref{eq:cond_energy}:
\begin{align}
    \frac{du}{dt} = \frac{1}{\rho} \nabla \cdot (\kappa \nabla T). 
\end{align}
However, we note that in astrophysical plasmas the conductivity is not constant but shows a dependence on the temperature. In the Spitzer limit this is given by 
\begin{align}
    \kappa_\mathrm{sp} = 1.31 n_{e} \lambda_{e} k_{B} \left(\frac{k_{B} T_{e}}{m_{e}} \right)^{1/2}, 
\end{align}
with the electron density n$_{e}$, its temperature T$_{e}$ and its mass m$_{e}$. However, $n_{e} \lambda_{e}$ is only a function of the temperature T$_{e}$ and the Coulomb-logarithm following
\begin{align}
    n_e \lambda_e = \frac{3^{3/2} (k_B T_e)^2}{4 \pi^{1/2} e^4 \ln \Lambda}.
\end{align}
In the case of a constant value for the Coulomb-logarithm the Spitzer conductivity remains strongly temperature dependent
\begin{align}
    \kappa_\mathrm{sp} = 8.2 \cdot 10^{20} \left(\frac{k_\mathrm{B}T}{10 \text{keV}}\right)^{5/2} \frac{\text{erg}}{\text{cm s keV}}.
\end{align}
However, the behaviour changes for very low plasma densities where the scale length of the temperature gradient is of order the electron mean free path or even smaller. In this regime the heat flux saturates and no longer increases even if the temperature gradient rises. The saturation flux is given by
\begin{align}
    j_\mathrm{sat} \approx 0.4 n_{e} k_{B} T \left( \frac{2 k_{B}T}{\pi m_{e}}\right)^{1/2}.
\end{align}
In this case $\kappa$ is modified to an effective $\kappa_\mathrm{eff}$ to smoothly change between the Spitzer and the saturated regime in the following manner
\begin{align}
    \kappa_\mathrm{eff} = \frac{\kappa_\mathrm{sp}}{1+4.2 \lambda_{e}/l_{T}},
\end{align}
where l$_{T}$ is the length scale of the temperature gradient given as $T/|\nabla T|$. While the saturation limit is mostly relevant on galaxy-cluster scales and not in the ISM, we carried out some test simulations without the heat flux limit and even without a temperature dependence of $\kappa$ itself. For isolated events we do not find large differences and it seems to be more important to have a heat diffusion term in the first place. The reason for that is, that we cannot resolve the process of thermal conduction within the bubble or the shell respectively. However, what we can resolve is the interface between bubble and shell. Therefore, even a constant value for $\kappa$ leads to similar results (as long as it is physically motivated), as $\kappa$ at the interface of bubble and shell stays (roughly) constant. 

\subsection{Chemistry model \label{sec:chemical_model}}

We adopt a non-equilibrium chemical model following the implementation of the 'SILCC' project \citep{Walch2015, Girichidis2016, Gatto2017, Peters2017}. This chemical model is based on \citet{Nelson1997}, \citet{Glover2007a}, \citet{Glover2007b} and  \citet{Glover2012}. We track six individual species, H$_{2}$, H$^{+}$, CO, H, C$^{+}$ and O as well as free electrons and assume that silicon is present in the simulation as Si$^{+}$. Carbon is present either in single ionised from as C$^{+}$ or CO. Oxygen is present as O. H$_{2}$ and H$^{+}$ undergo several reactions, which are summarised in table 1 of \citet{Micic2012}. The formation of H$_{2}$ is a dust catalysed reaction in which a dust grain captures two hydrogen atoms and the reaction becomes possible on dust surfaces. Most important for the destruction of H$_{2}$ is photo-dissociation due to the inter-stellar radiation field (ISRF), ionisation due to cosmic rays and via collisional dissociation (collisions with other species). The main channels for the formation of H$^{+}$ are collisional excitation with cosmic rays and free electrons. The destruction of H$^{+}$ is driven by recombination processes with free electrons. In the case of thermal conduction we directly assume the amount of free electrons that is predicted from solving the non-equilibrium rate equations to perform our computation of heat transfer. 

\subsection{Cooling and heating processes \label{sec:cooling_processes}}

The non-equilibrium cooling rates take into account the local density, temperature and the local chemical abundances. There are six major non-equilibrium cooling processes that are relevant for our studies, fine structure line cooling of atoms and ions (C$^{+}$, Si$^{+}$, O), vibration and rotation line cooling of molecules (H$_{2}$, CO), Lyman-alpha cooling of hydrogen, collisional dissociation of H$_{2}$, collisional ionisation of hydrogen and recombination of H$^{+}$ (both in gas phase and condensed on dust grains).
The main sources for heating are given by photo-electric heating from dust grains and polycyclic aromatic hydrocarbonates (PAHs), cosmic ray ionisation, photo-dissociation of H$_{2}$, UV-pumping of H$_{2}$ and formation of H$_{2}$.
For high temperatures ($T > 3 \cdot10^4$ K) we adopt the cooling function by \citet{Wiersma2009}. It assumes that the ISM is in equilibrium where the cooling due to collisional ionisation is balanced by the heating of the cosmic UV-radiation \citep{Haardt2001}. Further, the absolute cooling rate is determined by following the model of \citet{Aumer2013} and depends on the local gas density, gas temperature and includes metal line cooling from 11 species  (H, He, C, N, O, Ne, Mg, Si, S, Ca and Fe). 

\subsection{Shielding mechanisms from the interstellar radiation field}

For the hydrogen chemistry the ISRF photo dissociation rate R$_\mathrm{pd, H_2}$ acting on H$_2$ is affected by the shielding of dust and the self shielding of H$_{2}$. Dust has the ability to absorb photons of the ISRF field and re-emit in the infrared while absorbing the energy difference. Self-shielding by molecular hydrogen is more complicated. The process normally refers to a mechanism where the photo excitation transitions become optically thick and shield each other if UV radiation is coming from a single direction. However, in the optical thick limit (high column densities) this process is prevented by the line broadening of the Lyman and Werner bands \citep[e.g.][]{Gnedin2014}. In this regime the self shielding by dust becomes the more important process. We include both processes, the self-shielding by dust and molecular hydrogen. This is implemented with the \textsc{TreeCol}-algorithm \citep{Clark2012, Hu2016}.

\subsection{Numerical resolution and performance}
\label{sec:MFMres}

Although the single particles have very small masses ranging from 0.1 M$_{\odot}$ to 100 M$_{\odot}$ this is not the effective resolution of the simulations. For methods that follow an SPH-like volume distribution the relevant spatial resolution is the size of the kernel or to be more specific the radius of compact support. For the SPH runs we use a Wendland C4 kernel with $100$ neighbours while we use a cubic spline with $32$ neighbours for the MFM runs to capture the fluid properties. The effective spatial and mass resolution is higher for the runs with MFM as the radius of compact support is smaller for the cubic spline and it can operate with a smaller number of neighbours. The resulting mass resolution can be calculated via 
\begin{align}
  m_\mathrm{resolved} = N_\mathrm{ngb} m_\mathrm{gas} \left(\frac{h}{H}\right)^3,
\end{align}
where $h$ is the smoothing length and $H$ is the radius of compact support of the respective kernel.
For the cubic spline we find the relation $h=0.55H$ and for the Wendland C4 kernel $h=0.45H$. Therefore, the resolution for MFM is better by about a factor of $2$. 
While the MFM method requires more memory than SPH methods (a factor of $1.5$) it is overall faster in solving the equations of hydrodynamics compared to our state-of-the-art SPH solver (for the same number of neighbours).

\begin{figure}
        \includegraphics[scale=0.5]{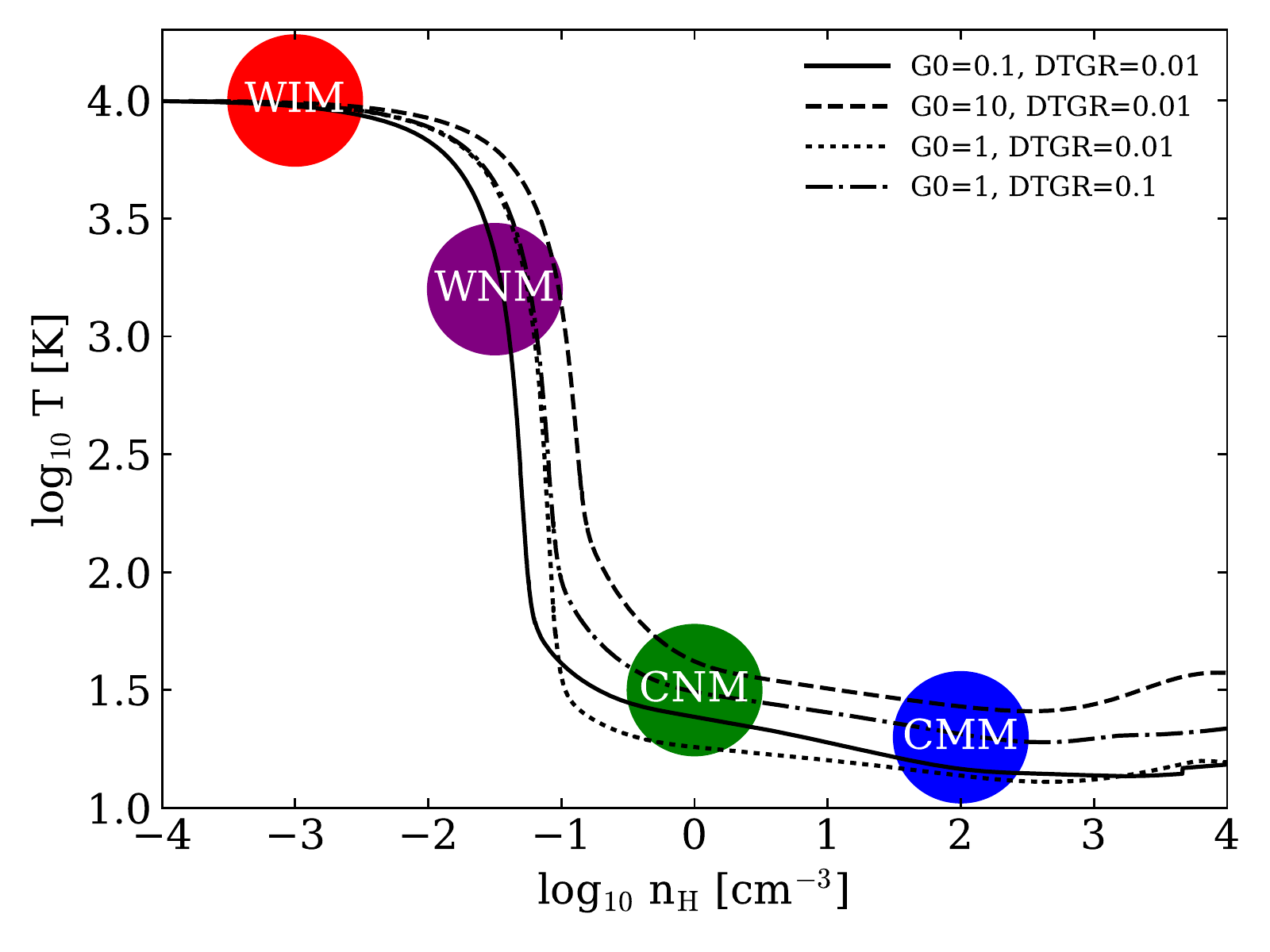}
        \includegraphics[scale=0.5]{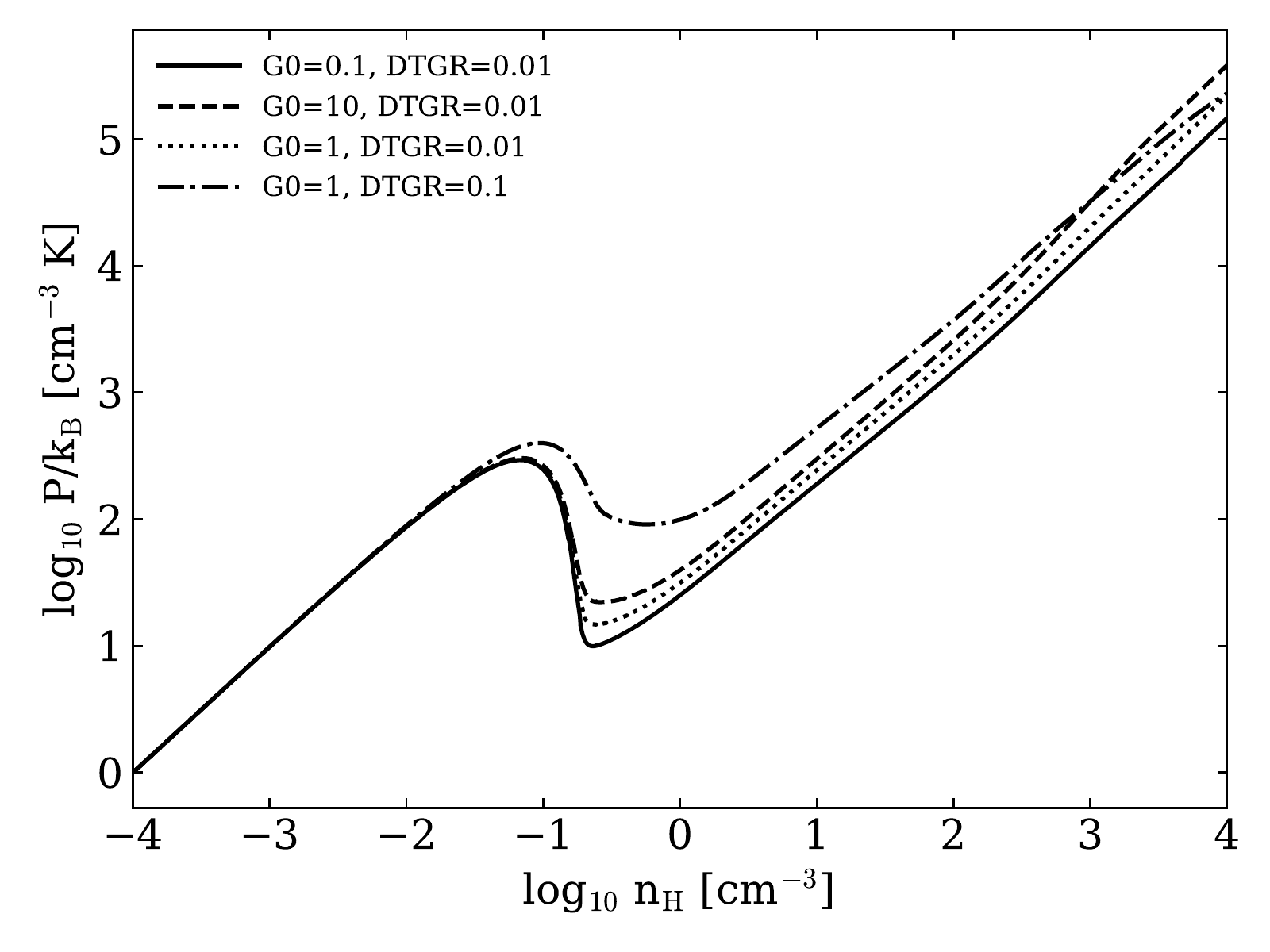}
        \caption{Equilibrium phase diagrams for temperature (top) and pressure (bottom) for a metallicity of 0.1 $Z_{\odot}$. We show the curves for different values of $G_{0}$ and dust-to-gas ratios (DTGR). These variations only affect the high density and low temperature regimes. The four different ISM regimes, warm ionised medium (WIM), warm neutral medium (WNM), cold neutral medium (CNM) and cold molecular medium (CMM) for the SN blast wave simulations are indicated by the coloured circles. The thermally unstable regime at densities $10^{-2}$ to 10$^{-1}$ cm$^{-3}$ can be identified in the bottom panel.}
        \label{fig:equilibrium_cooling_curve}
\end{figure}

\begin{table*}
    \centering
    \caption{Simulation parameters for supernovae in homogeneous ambient media}
    \label{tab:Table1}
    \begin{tabular}{lcccccccc}
    \hline
    \hline
      Simulation Name & ambient n [cm$^{-3}$] & ambient T [K] &  metallicity [Z$_{\odot}$] & chemistry & m$_\mathrm{gas}$ [M$_{\odot}$]  & solver & SN-feedback \\ 
      & & & & & & & resolved\\ \hline
      \multicolumn{8}{c}{\textbf{Cold Molecular Medium}}\\ \hline
      \textbf{CMM-L1} & $100$ & $10$ & $0.1$ & $100$\% H$_{2}$ & $100$ & MFMH & \textcolor{red}{\xmark} \\
      \textbf{CMM-L2} & $100$ & $10$ & $0.1$ & $100$\% H$_{2}$ & $10$  & MFM & $\bullet$\\
      \textbf{CMM-L3} & $100$ & $10$ & $0.1$ & $100$\% H$_{2}$ & $1$  & MFM & \textcolor{green}{\cmark}\\
      \textbf{CMM-L4} & $100$ & $10$ & $0.1$ & $100$\% H$_{2}$ & $0.1$  & MFM, SPH & \textcolor{green}{\cmark}\\
      \textbf{CMM-cond-L4} & $100$ & $10$ & $0.1$ & $100$\% H$_{2}$ & $0.1$  & SPH & \textcolor{green}{\cmark}\\
      
      \textbf{CMM-Z1-L1} & $100$ & $10$ & $1$ & $100$\% H$_{2}$ & $100$   & MFM & \textcolor{red}{\xmark}\\
      \textbf{CMM-Z1-L2} & $100$ & $10$ & $1$ & $100$\% H$_{2}$ & $10$   & MFM & $\bullet$\\
      \textbf{CMM-Z1-L3} & $100$ & $10$ & $1$ & $100$\% H$_{2}$ & $1$  & MFM & \textcolor{green}{\cmark} \\
      \textbf{CMM-Z1-L4} & $100$ & $10$ & $1$ & $100$\% H$_{2}$ & $0.1$   & MFM & \textcolor{green}{\cmark}\\
      
      \textbf{CMM-Z001-L1} & $100$ & $10$ & $0.01$ & $100$\% H$_{2}$ & $100$   & MFM & \textcolor{red}{\xmark} \\
      \textbf{CMM-Z001-L2} & $100$ & $10$ & $0.01$ & $100$\% H$_{2}$ & $10$   & MFM & $\bullet$\\
      \textbf{CMM-Z001-L3} & $100$ & $10$ & $0.01$ & $100$\% H$_{2}$ & $1$   & MFM & \textcolor{green}{\cmark}\\
      \textbf{CMM-Z001-L4} & $100$ & $10$ & $0.01$ & $100$\% H$_{2}$ & $0.1$  & MFM & \textcolor{green}{\cmark}\\
      \hline
     
     \multicolumn{8}{c}{\textbf{Cold Neutral Medium}}\\ \hline
     \textbf{CNM-n10-L1} & $10$ & $100$ & $0.1$ & $100$\% H & $100$ & MFM & \textcolor{red}{\xmark}\\
      \textbf{CNM-n10-L2} & $10$ & $100$ & $0.1$ & $100$\% H & $10$ & MFM & $\bullet$\\
      \textbf{CNM-n10-L3} & $10$ & $100$ & $0.1$ & $100$\% H & $1.0$ & MFM & \textcolor{green}{\cmark}\\
      \textbf{CNM-n10-L4} & $10$ & $100$ & $0.1$ & $100$\% H & $0.1$ & MFM, SPH & \textcolor{green}{\cmark}\\
      \textbf{CNM-n10-cond-L4} & $10$ & $100$ & $0.1$ & $100$\% H & $0.1$  & SPH & \textcolor{green}{\cmark}\\
      
      \textbf{CNM-L1} & $1$ & $100$ & $0.1$ & $100$\% H & $100$ & MFM, SPH & \textcolor{red}{\xmark}\\
      \textbf{CNM-L2} & $1$ & $100$ & $0.1$ & $100$\% H & $10$ & MFM, SPH & $\bullet$\\
      \textbf{CNM-L3} & $1$ & $100$ & $0.1$ & $100$\% H & $1.0$ & MFM, SPH & \textcolor{green}{\cmark}\\
      \textbf{CNM-L4} & $1$ & $100$ & $0.1$ & $100$\% H & $0.1$ & MFM, SPH & \textcolor{green}{\cmark} \\
      \textbf{CNM-cond-L4} & $1$ & $100$ & $0.1$ & $100$\% H & $0.1$ & SPH & \textcolor{green}{\cmark}\\
      
      \textbf{CNM-Z1-L1} & $1$ & $100$ & $1$ & $100$\% H & $100$ &  MFM, SPH & \textcolor{red}{\xmark}\\
      \textbf{CNM-Z1-L2} & $1$ & $100$ & $1$ & $100$\% H & $10$  &   MFM, SPH & $\bullet$\\
      \textbf{CNM-Z1-L3} & $1$ & $100$ & $1$ & $100$\% H & $1.0$  &    MFM, SPH & \textcolor{green}{\cmark}\\
      \textbf{CNM-Z1-L4} & $1$ & $100$ & $1$ & $100$\% H & $0.1$   & MFM, SPH & \textcolor{green}{\cmark}\\
      
      \textbf{CNM-Z001-L1} & $1$ & $100$ & $0.01$ & $100$\% H & $100$  & MFM, SPH & \textcolor{red}{\xmark}\\
      \textbf{CNM-Z001-L2} & $1$ & $100$ & $0.01$ & $100$\% H & $10$   & MFM, SPH & $\bullet$\\
      \textbf{CNM-Z001-L3} & $1$ & $100$ & $0.01$ & $100$\% H & $1.0$  & MFM, SPH & \textcolor{green}{\cmark}\\
      \textbf{CNM-Z001-L4} & $1$ & $100$ & $0.01$ & $100$\% H & $0.1$  & MFM, SPH & \textcolor{green}{\cmark}\\
      \hline
     
     \multicolumn{8}{c}{\textbf{Warm Neutral Medium}}\\ \hline
      \textbf{WNM-L1} & $0.1$ & $4000$ & $0.1$ & $100$\% H & $100$   & MFM & \textcolor{red}{\xmark}\\
      \textbf{WNM-L2} & $0.1$ & $4000$ & $0.1$ & $100$\% H & $10$   & MFM & \textcolor{green}{\cmark}\\
      \textbf{WNM-L3} & $0.1$ & $4000$ & $0.1$ & $100$\% H & $1.0$   & MFM & \textcolor{green}{\cmark}\\
      \textbf{WNM-L4} & $0.1$ & $4000$ & $0.1$ & $100$\% H & $0.1$   & MFM, SPH & \textcolor{green}{\cmark}\\
      \textbf{WNM-cond-L4} & $0.1$ & $4000$ & $0.1$ & $100$\% H & $0.1$ & SPH & \textcolor{green}{\cmark}\\
      
      \textbf{WNM-Z1-L1} & $0.1$ & $4000$ & $1$ & $100$\% H & $100$   & MFM, SPH & \textcolor{red}{\xmark}\\
      \textbf{WNM-Z1-L2} & $0.1$ & $4000$ & $1$ & $100$\% H & $10$   & MFM & \textcolor{green}{\cmark}\\
      \textbf{WNM-Z1-L3} & $0.1$ & $4000$ & $1$ & $100$\% H & $1.0$   & MFM & \textcolor{green}{\cmark}\\
      \textbf{WNM-Z1-L4} & $0.1$ & $4000$ & $1$ & $100$\% H & $0.1$   & MFM & \textcolor{green}{\cmark}\\
      
      \textbf{WNM-Z001-L1} & $0.1$ & $4000$ & $0.01$ & $100$\% H & $100$   & MFM & \textcolor{red}{\xmark}\\
      \textbf{WNM-Z001-L2} & $0.1$ & $4000$ & $0.01$ & $100$\% H & $10$   & MFM & \textcolor{green}{\cmark}\\
      \textbf{WNM-Z001-L3} & $0.1$ & $4000$ & $0.01$ & $100$\% H & $1.0$   & MFM & \textcolor{green}{\cmark}\\
      \textbf{WNM-Z001-L4} & $0.1$ & $4000$ & $0.01$ & $100$\% H & $0.1$   & MFM, SPH & \textcolor{green}{\cmark}\\
      \hline
      \multicolumn{8}{c}{\textbf{Warm Ionised Medium}}\\ \hline
      \textbf{WIM-L1} & $0.01$ & $8000$ & $0.1$ & $100$\% H$^{+}$ & $100$   & MFM & \textcolor{red}{\xmark}\\
      \textbf{WIM-L2} & $0.01$ & $8000$ & $0.1$ & $100$\% H$^{+}$ & $10$   & MFM & \textcolor{green}{\cmark}\\
      \textbf{WIM-L3} & $0.01$ & $8000$ & $0.1$ & $100$\% H$^{+}$ & $1.0$   & MFM & \textcolor{green}{\cmark}\\
      \textbf{WIM-L4} & $0.01$ & $8000$ & $0.1$ & $100$\% H$^{+}$ & $0.1$   & MFM, SPH & \textcolor{green}{\cmark}\\
      \textbf{WIM-cond-L4} & $0.01$ & $8000$ & $0.1$ & $100$\% H$^{+}$ & $0.1$  & SPH & \textcolor{green}{\cmark}\\
      
       \hline
      \hline
    \end{tabular}
  \end{table*}
  
\begin{table*}
    \centering
    \label{tab:Table2}
    \begin{tabular}{lcccccccc}
     \hline
     \hline
      Simulation Name & ambient n [cm$^{-3}$] & ambient T [K] &  metallicity [Z$_{\odot}$] & chemistry & m$_\mathrm{gas}$ [M$_{\odot}$]  & solver & SN-feedback \\ 
      & & & & & & & resolved\\ \hline
     \multicolumn{8}{c}{\textbf{Warm Ionised Medium (continued)}}\\ \hline
      
      \textbf{WIM-Z1-L1} & $0.01$ & $8000$ & $1$ & $100$\% H$^{+}$ & $100$  &  MFM & \textcolor{red}{\xmark}\\
      \textbf{WIM-Z1-L2} & $0.01$ & $8000$ & $1$ & $100$\% H$^{+}$ & $10$   & MFM & \textcolor{green}{\cmark}\\
      \textbf{WIM-Z1-L3} & $0.01$ & $8000$ & $1$ & $100$\% H$^{+}$ & $1.0$ &  MFM & \textcolor{green}{\cmark}\\
      \textbf{WIM-Z1-L4} & $0.01$ & $8000$ & $1$ & $100$\% H$^{+}$ & $0.1$  &  MFM & \textcolor{green}{\cmark}\\
      
      \textbf{WIM-Z001-L1} & $0.01$ & $8000$ & $0.01$ & $100$\% H$^{+}$ & $100$  &  MFM & \textcolor{red}{\xmark}\\
      \textbf{WIM-Z001-L2} & $0.01$ & $8000$ & $0.01$ & $100$\% H$^{+}$ & $10$  & MFM & \textcolor{green}{\cmark}\\
      \textbf{WIM-Z001-L3} & $0.01$ & $8000$ & $0.01$ & $100$\% H$^{+}$ & $1.0$  &  MFM & \textcolor{green}{\cmark}\\
      \textbf{WIM-Z001-L4} & $0.01$ & $8000$ & $0.01$ & $100$\% H$^{+}$ & $0.1$  &  MFM & \textcolor{green}{\cmark}\\

      \textbf{WIM-n0001-L1} & $0.001$ & $10000$ & $0.1$ & $100$\% H$^{+}$ & $100$   & MFM, SPH & \textcolor{red}{\xmark}\\
      \textbf{WIM-n0001-L2} & $0.001$ & $10000$ & $0.1$ & $100$\% H$^{+}$ & $10$   & MFM & \textcolor{green}{\cmark}\\
      \textbf{WIM-n0001-L3} & $0.001$ & $10000$ & $0.1$ & $100$\% H$^{+}$ & $1.0$   & MFM & \textcolor{green}{\cmark}\\
      \textbf{WIM-n0001-L4} & $0.001$ & $10000$ & $0.1$ & $100$\% H$^{+}$ & $0.1$   & MFM & \textcolor{green}{\cmark}\\
      \textbf{WIM-n0001-cond-L4} & $0.001$ & $10000$ & $0.1$ & $100$\% H$^{+}$ & $0.1$  & SPH & \textcolor{green}{\cmark}\\
      
      \textbf{WIM-Z1-L4} & $0.001$ & $10000$ & $1$ & $100$\% H$^{+}$ & $0.1$  &  MFM, SPH & \textcolor{green}{\cmark}\\
      
      \textbf{WIM-Z001-L4} & $0.001$ & $10000$ & $0.01$ & $100$\% H$^{+}$ & $0.1$  &  MFM, SPH & \textcolor{green}{\cmark}\\
      
      \hline
      \hline
    \end{tabular}
  \end{table*}

\begin{figure*}
        \centering
        \includegraphics[scale=0.5]{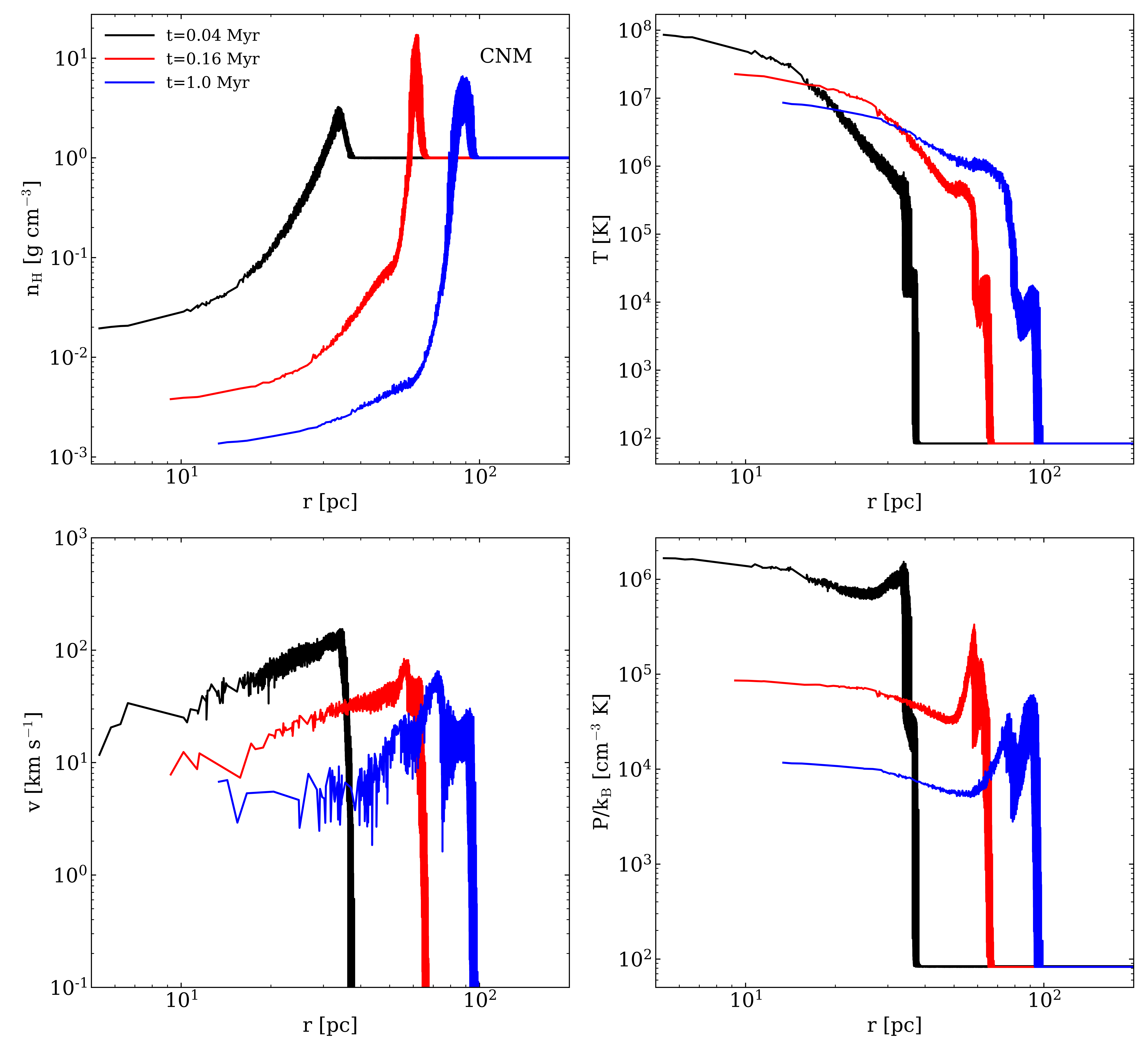}
        \caption{Hydrogen number density (top left), temperature (top right), velocity (bottom left) and pressure (bottom right) for the simulation CNM-L4 as a function of the radius. We display the results at three different points in time that represent different phases of the blast wave. The black lines indicate $t=0.04$ Myr where the shock is at the end of the energy conserving ST-phase. The red lines show $t=0.14$ Myr. In this regime the cooling time is shorter than the total time of the simulation (roughly $0.08$ Myr) and the system is dominated by cooling. The blue lines show the properties of the shock at the end of the simulation when the shock merges with the ambient medium. All properties of the shock are resolved in each investigated regime of the shock. \textit{Top Left:} In the ST-phase (black line) the shock is resolved in the adiabatic regime. In the cooling dominated regime (red line) the maximum density exceeds the expected adiabatic solution by a factor $3.5$. At the end of the simulation (blue line) the density decreases again and the shock starts to merge with the ambient medium. \textit{Top Right:} The temperatures in the post-shock region are very high (above $10^{8}$ K). When cooling dominates energy is radiated away and the temperature starts to decrease by an order of magnitude. \textit{Bottom Left:} In the beginning the shock moves very fast outwards (a few 100 km s$^{-1}$) but gets slower at later times. \textit{Bottom Left:} The pressure is roughly constant during the ST-phase of the shock but starts to decrease very fast after shell formation.}
        \label{fig:all_dens}
\end{figure*}

\begin{figure*}
        \centering
        \includegraphics[scale=0.5]{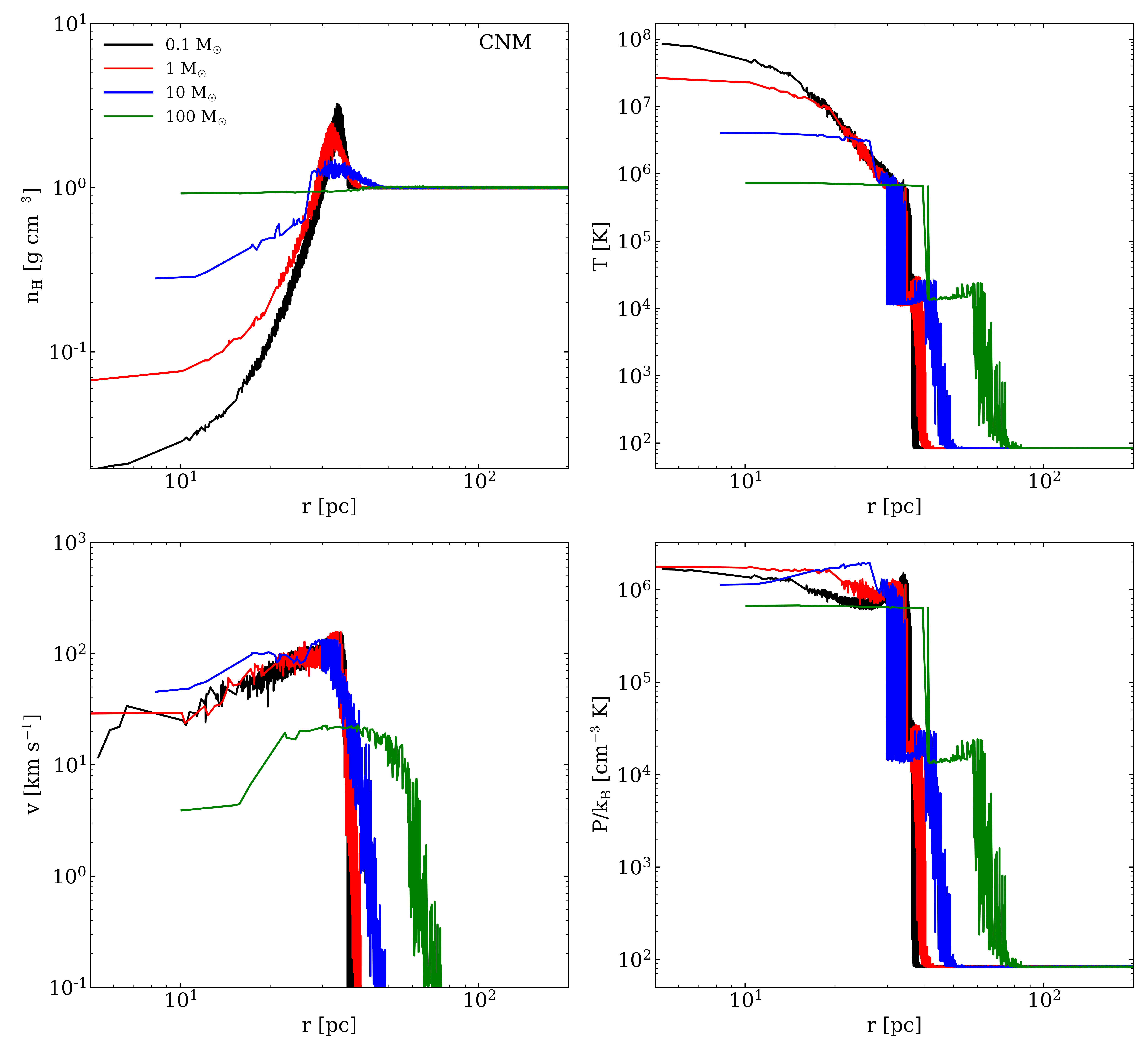}
        \caption{Different properties of the SN-remnant as a function of the radius in the ST-phase (t$=0.04$ Myr) for all mass resolutions in the CNM. The black lines show the $0.1$ M$_{\odot}$, the red lines the $1$ M$_{\odot}$ resolution, the blue lines the $10$ M$_{\odot}$ and the green lines the $100$ M$_{\odot}$ resolution. \textit{Top Left:} Density of the shock as a function of radius. With decreasing resolution the shock becomes less and less resolved. At the two highest mass resolutions we can detect the shock very well in the ST-phase. A the highest mass resolution we find the maximum density increased by a factor $3.5$ compared to the ambient density. While we can still detect the shock in the ST-phase for the $10$ M$_{\odot}$ resolution the shock in the density completely vanishes at the lowest resolution. \textit{Top Right:} Temperature structure of the shock. The three highest mass resolutions agree well apart from the maximum temperature within the bubble. At the lowest mass resolution, we find that the shock is at a different radial position which corresponds to the larger injection region in this regime. \textit{Bottom Left:} Velocity structure of the shock in the ST-phase. The three highest mass resolutions agree very well, while at the lowest resolution the velocity structure remains unresolved. \textit{Bottom Right:} Pressure in the remnant. For the three highest resolutions we find that the pressure is well resolved. For the lowest resolution the pressure in the bubble is already significantly lower, compared to the higher resolution runs.}
        \label{fig:all_dens_res}
\end{figure*}

\subsection{Initialisation and naming convention}
In our default runs we adopt a metallicity of $Z=0.1$ Z$_{\odot}$ with a constant the dust to gas ratio mass ratio of 0.01. The constant UV-background \citep{Haardt2012} has a normalisation factor of $G_{0}=1.7$ for the interstellar radiation field, which corresponds to the value of the solar neighbourhood.  
For resolution we refer to the gas particle mass. The four resolution levels are $100$ M$_{\odot}$ (L1), $10$ M$_{\odot}$ (L2), $1$ M$_{\odot}$ (L3) and $0.1$ M$_{\odot}$ (L4). The SN blast waves are simulated in the cold molecular medium (CMM) with $n=100$ cm$^{-3}$, the cold neutral medium (CNM) with $n=1$ cm$^{-3}$, the warm neutral medium (WMN) with $n=0.1$ cm$^{-3}$ and the warm ionised medium. Some runs have lower $Z=0.01$ $Z_{\odot}$ (Z-001) and and higher $Z=1 Z_{\odot}$ (Z-1) metallicities. A different number density compared to our defaults is indicated with n and the number density in units of cm$^{-3}$. Runs with isotropic thermal conduction are indicated by \textit{cond}.

\begin{figure*}
        \centering
        \includegraphics[scale=0.5]{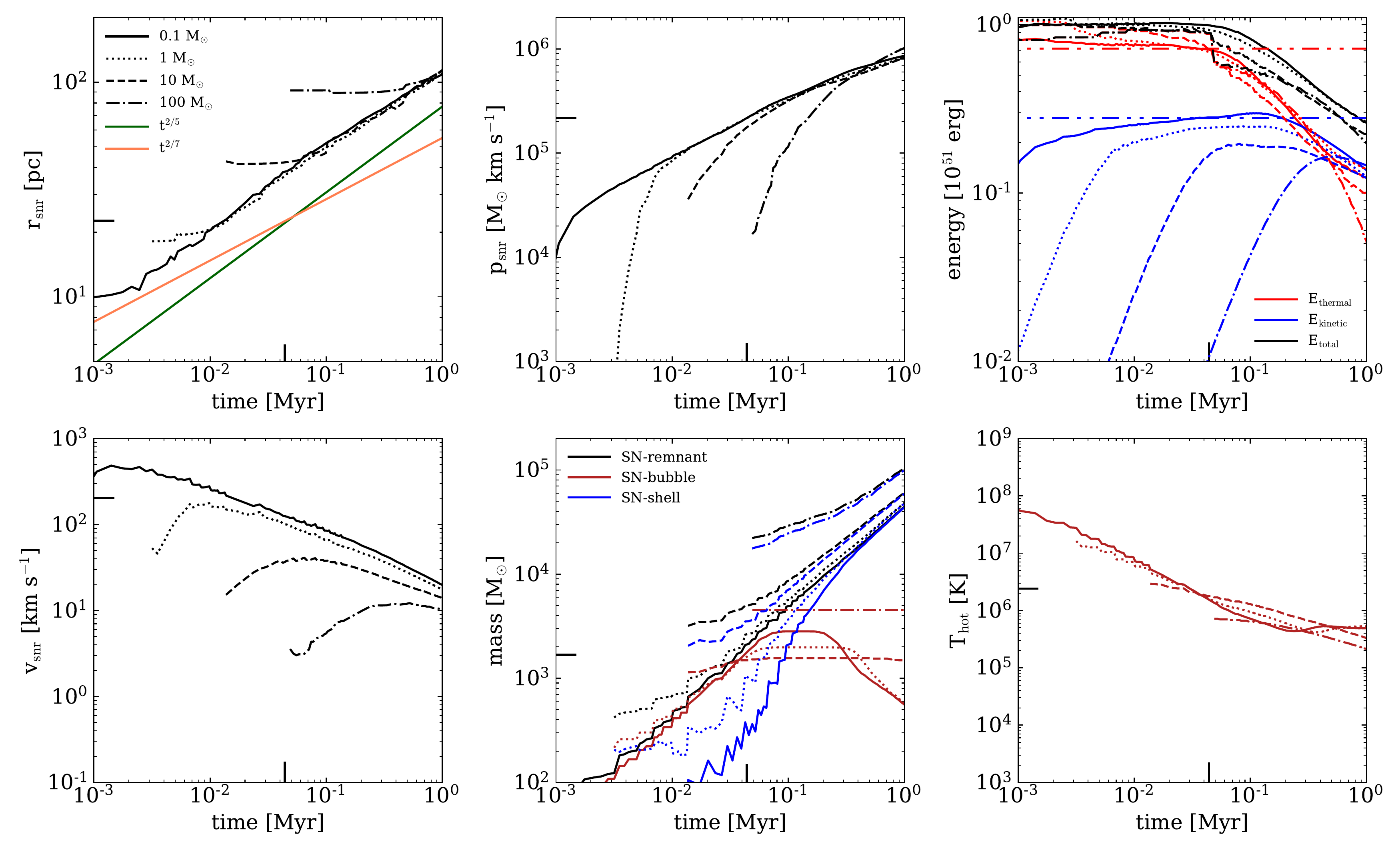}
        \caption{Evolution of supernova remnants in the CNM at four different resolution levels of $0.1$ M$_{\odot}$ (solid),  $1$ M$_{\odot}$ (dotted),  $10$ M$_{\odot}$ (dashed) and  $100$ M$_{\odot}$ (dashed-dotted). The black markers on the axis represent the physical values at the end of the ST-phase. \textit{Top Left:} Time evolution of the shell radius with the analytical values of $r \propto t^{2/5}$ (green) and $r \propto t^{2/7}$ (red) for Sedov phase and the momentum conserving phase, respectively. Our results converge for mass resolution higher than $10$ M$_{\odot}$. \textit{Top Middle:} The shell momentum as a function of time is relatively insensitive to resolution. \textit{Top right:} Total energy (black), kinetic energy (blue) and thermal energy (red) as a function of time. The total energy is conserved in the ST-phase with $28$ per cent kinetic and $72$ per cent thermal energy (dot-dot-dashed lines). The kinetic energy builds up during shell formation (we inject only thermal energy) and reaches the ST-solution only for the highest mass resolution. The  $10$ M$_{\odot}$ run only creates $\sim$ $20$ per cent kinetic energy. \textit{Bottom Left:} The shell velocity as a function of time strongly depends on resolution. At the end of the Sedov phase it ranges from $30$ km s$^{-1}$ to $200 km s{-1}$ for low and high resolution, respectively. \textit{Bottom Middle:} The shell mass a function of time is smaller at higher resolution. \textit{Bottom right:} Bubble temperature (red) as a function of time. At low mass resolution the bubble temperature and therefore its pressure remain unresolved. At mass resolution higher than 10 M$_{\odot}$ the bubble temperature at the Sedov time is captured. At 100 M$_{\odot}$ the temperature evolution is unresolved.}
        \label{fig:1cmregion}
\end{figure*}
\section{Individual supernova blast waves}
\label{sec:idealised}

We present a set of simulations of individual SN blast waves in six different environments at equilibrium conditions of the ISM for our cooling model. In the upper panel of Figure \ref{fig:equilibrium_cooling_curve} we show the temperature as a function of hydrogen number density for different values for a (constant) interstellar radiation field and the dust to gas ratio (DTGR). The coloured circles indicate the ISM conditions for WIM, WNM, CNM and CMM. The CMM represents regimes in dense molecular clouds ($n_{H} = 100$cm$^{-3}$, $T=10$ K, fully molecular). The CNM has higher temperatures and densities such that hydrogen is present in neutral atomic form ($n_{H} = 1$cm$^{-3}$, $T=100$ K, fully neutral). The WNM is warmer but still neutral ($n_{H} = 0.1$cm$^{-3}$, $T=8000$ K, fully neutral) and the WIM is fully ionised with long cooling times (n$_{H} = 0.001$cm$^{-3}$, T$=10000$ K, fully ionised). The pressure vs. density plot in the bottom panel highlights the pressure-unstable regime between $10^{-2}$ and $10^{-1}$ cm$^{-3}$. We initialise each supernova event by a kernel-weighted injection of $10^{51}$ erg of thermal energy into all kernel particles. For the MFM-solver we hereby use 32 particles and cubic-spline (M4) kernel. For the respective SPH runs we use a Wendland C4 kernel with 100 neighbours. Experience has shown that MFM with a cubic spline and $32$ neighbours shows the same quality of capturing fluid flows as the SPH reference runs with 100 neighbours and a Wendland C4 kernel. In the case of SPH the higher neighbour number is needed to suppress the intrinsic noise of particles methods by smoothing over the fluid field. However, the well known and investigated pairing-instability makes it than necessary to use a higher order kernel with positive Fourier transformation. The initial phase of free expansion is very short (only a few $100$ yr) and is not modelled here. The initial conditions are summarised in Table \ref{tab:Table1}.
\begin{figure}
        \includegraphics[scale=0.5]{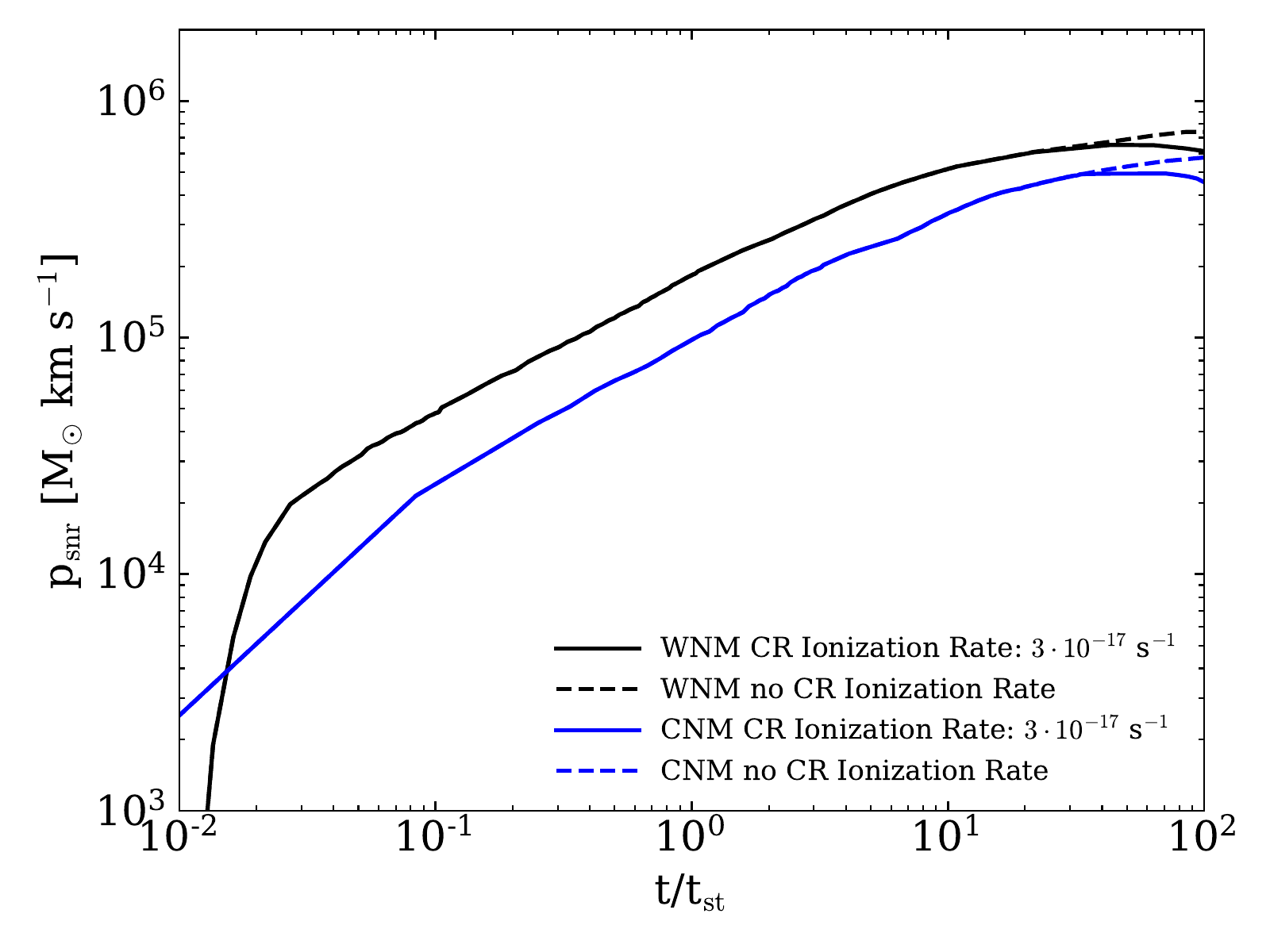}
        \caption{Momentum evolution normalised to the end of the ST-phase as defined in section \ref{sec:idealised} for two remnants with and without cosmic-ray ionisation heating included in the cooling function in the WNM and the CMM. The momentum conserving snow-plough phase is reached later and at higher shell momentum if cosmic-ray ionisation heating is not included. This is due to a lower ambient equilibrium temperature and a longer PDS phase. The pressurised shell is evolving into the lower pressure ambient medium, by 20 per cent, and delays saturation of momentum. This happens slightly earlier at lower ambient densities as the ambient temperature is higher than in the higher density regime.}
        \label{fig:momentum_conserving}
\end{figure}
\begin{figure*}
        \centering
        \includegraphics[scale=0.65]{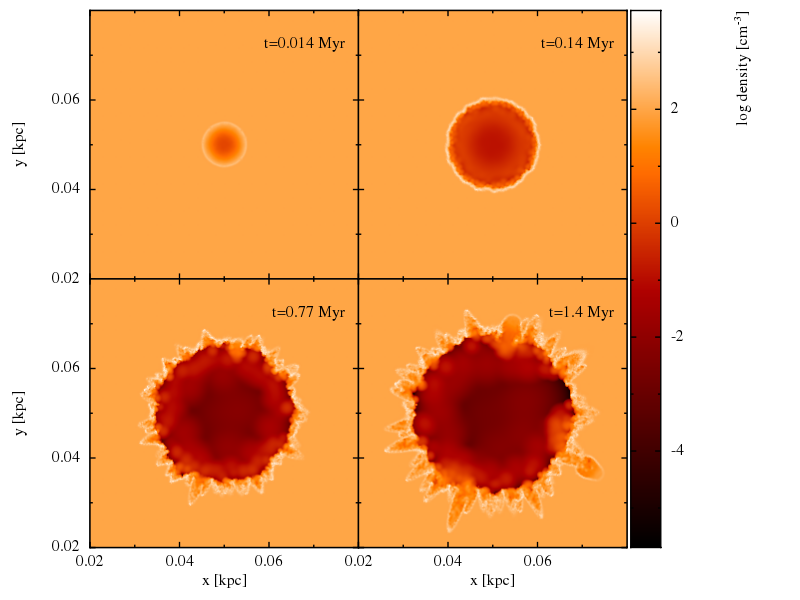}
        \caption{Time evolution for four different point in time of a supernova-feedback event in the cold molecular medium. Initially, in the ST-phase we can observe a spherical remnant (top left). Once the thin shell has formed (top right), the bubble supports the evolution of the shell until it they reach pressure equilibrium and the remnant moves outwards with constant velocity. At later times, the shell starts to become unstable due to Ryu-Vishniac-instabilities, that start to grow after roughly 100 ST-times (bottom left), until they dominate the structure in the outer parts of the remnant (bottom right).}
        \label{fig:blastwave_evo}
\end{figure*} 
\begin{figure*}
        \centering
        \includegraphics[scale=0.5]{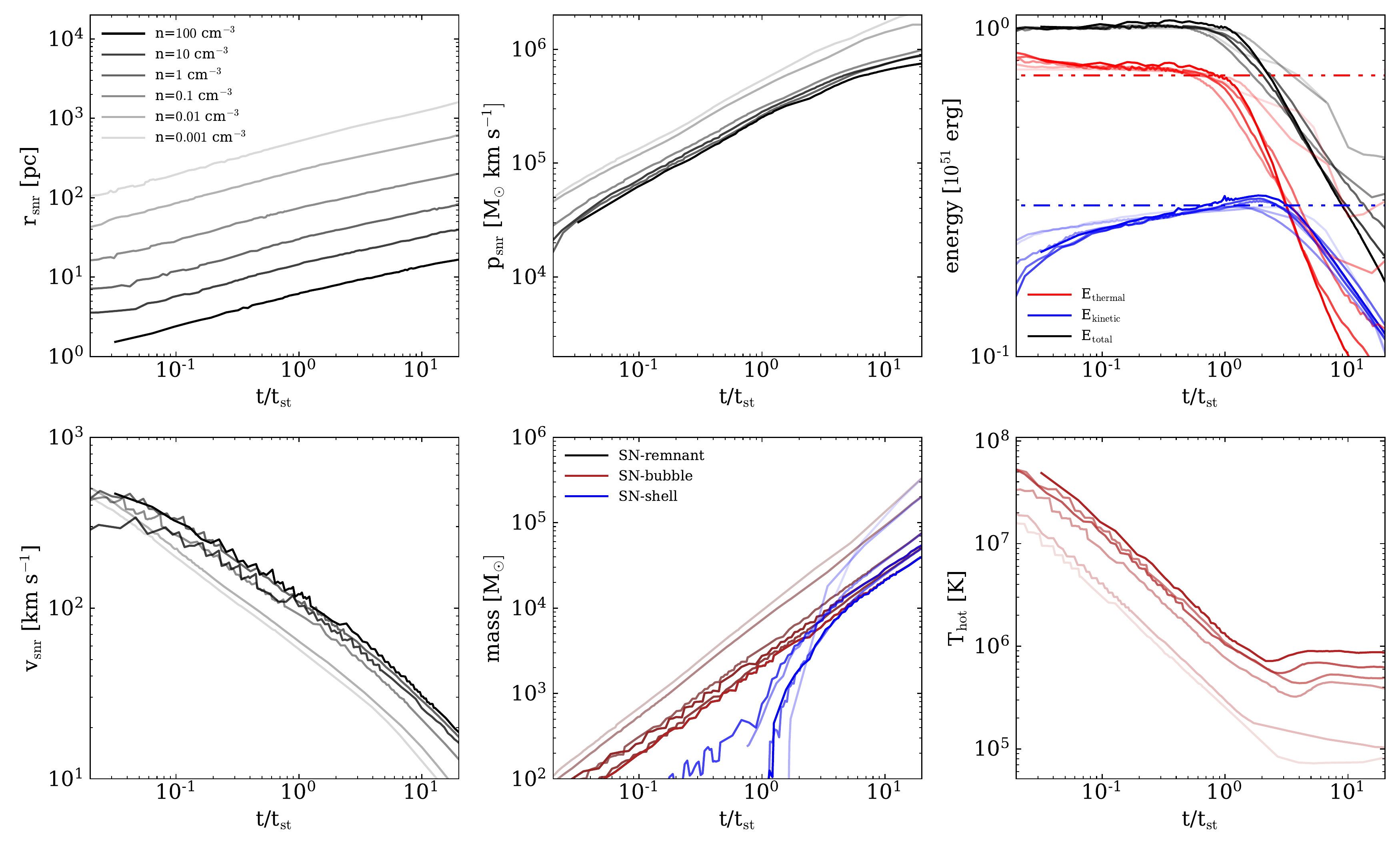}
        \caption{Comparison of physical properties of SN-blastwaves for different ambient densities for our highest all our highest resolution runs. For all runs the ST-phase is well resolved at our highest mass resolution. For better comparison we normalised the time axis to the respective ST-time that we derive in Figure \ref{fig:sedov_time}. \textit{Top Right:} In the high ambient media the radii become significantly smaller. While at the highest ambient medium the remnant has a size of a few parsec at the end of the ST-phase, we find that in the lowest ambient medium the remnant has a size of roughly $700$ parsec at the end of the ST-phase. \textit{Top Middle:} The remnants in ambient media with the highest density generate most momentum within the ISM. These remnants have the longest cooling times but fill the highest volume and therefore generate most momentum at the end of the ST-phase \textit{Top Right:} Energy in the shocked region split up in kinetic energy (blue), thermal energy (red) and total energy (black). In all environments the evolution towards the ST-phase is given and we find that $28$ per cent of the energy is deposited in kinetic energy while $72$ per cent is in thermal energy before cooling takes over and thermal energy is radiated away after the end of the ST-phase. \textit{Bottom Left:} In the beginning the remnants in the lowest density environments are the fastest, while at the end of the ST-phase they are the slowest. \textit{Bottom Middle:} The remnants with the highest mass at the end of ST-phase are those in the low density environments. Therefore, they have the highest momenta at the end of the ST-phase but also the lowest velocity. \textit{Bottom Right:} The temperature in the bubble decreases with decreasing ambient density. This phase is important because it generates the pressure in the bubble which makes it possible to drive outflows.}
        \label{fig:various_densities}
\end{figure*} 
\begin{figure}
        \includegraphics[scale=0.5]{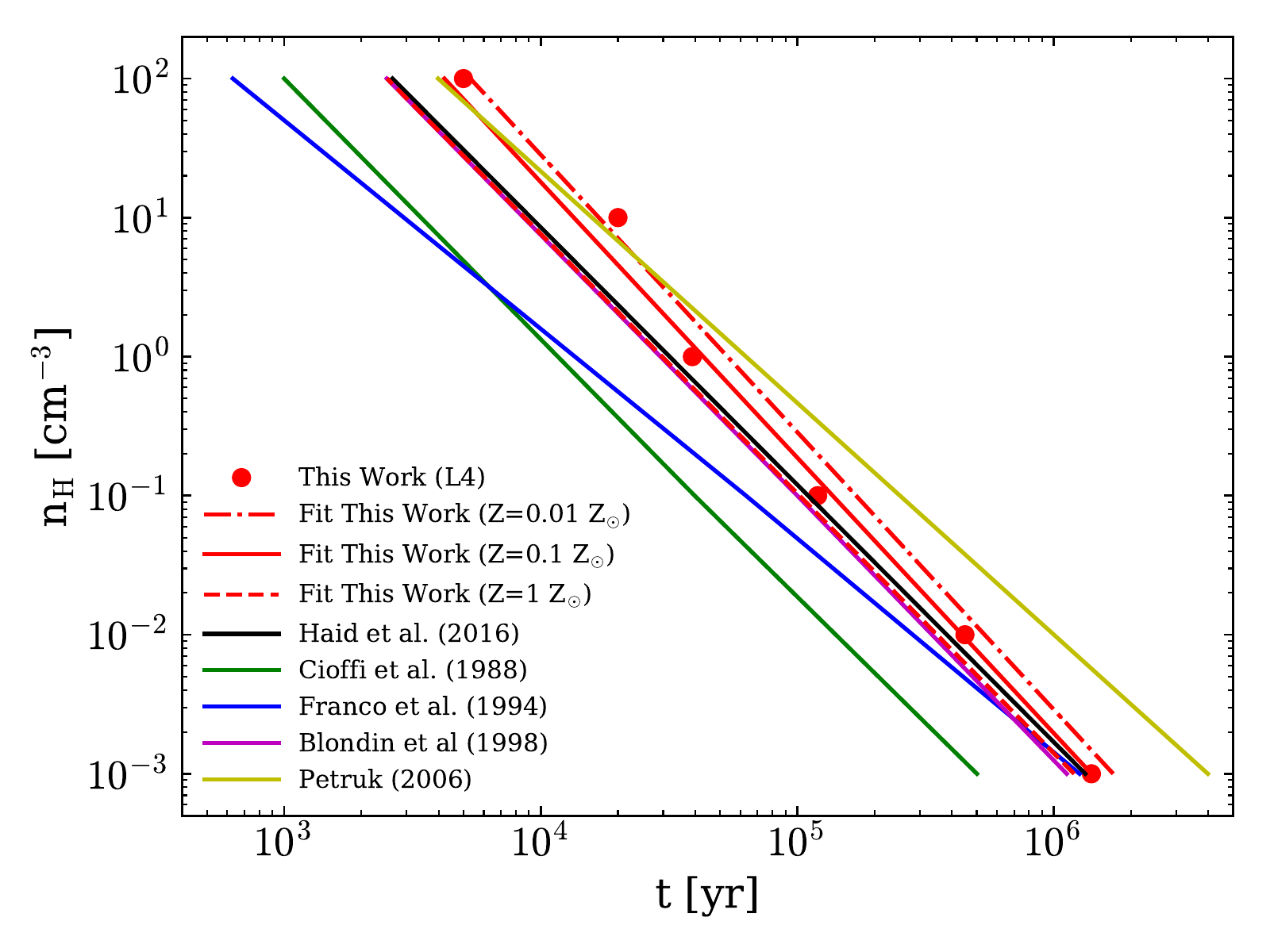}
        \caption{Shell formation times for different ambient densities at our highest resolution level. We show the data for our highest resolution run (red dots) for a metallicity of $Z=0.1$ $Z_{\odot}$ (red solid line). For the two other metallicities we tested we only show the fitting function that we obtain in the $Z=1$ $Z_{\odot}$ regime (red dashed line) and in the $Z=0.01$ $Z_{\odot}$ regime (red dashed-dotted line).  We find reasonable agreements with the fits provided by \citet{Blondin1998}, \citet{Haid2016}, \citet{Cioffi1988}, \citet{Franco1994} and \citet{Petruk2006}. Note that all authors use slightly different cooling functions and some results are from (semi-)analytical computations.}
        \label{fig:sedov_time}
\end{figure}

\subsection{Blast wave evolution}

In Figure \ref{fig:all_dens} we show the radial profiles of the hydrogen number density (top left), the temperature (top right), the relative velocity of the shock (bottom left) and the thermal pressure (bottom right) for a blast wave in the CNM at the highest resolution of $0.1$ M$_{\odot}$, which implies a spatial resolution of around $1$ pc. The different colours show the system at time $t=0.04$ Myr (black) during the ST-phase, at $t=0.16$ Myr (red) after the formation of a dense cooling shell, and at $t=1.0$ Myr (blue) when the remnant starts to merge with the ISM. In the ST-phase the maximum value of the density is around $3.53$ cm$^{-3}$. In the case of the pure adiabatic solution we would expect a maximum density of $4$ cm$^{-3}$ under constraint of the Rankine-Hugoniot jump-conditions with an adiabatic index of $5/3$ which is adopted in all our simulations.After the end of the ST-phase shell remnant has formed a thin, cooling dominated shell with a maximum density of roughly $20$ times the initial ambient density. At the end of the simulation the amplitude of the density decreases again and the remnant starts to merge with the ambient ISM. In all three stages the temperature declines as a function of the radius. After shell we can observe a bump in the temperature distribution at around $10^4$ K which can be identified as the dense cooling shell of the remnant. This shell grows over time as more and more gas from the initially hot bubble starts to cool. The velocities within the ST-phase are high which can lead to a velocity dispersion of roughly $10$ km s$^{-1}$ that is observed within the ISM. At later stages the remnant slows down until it finally reaches the point where it merges with the ISM. Initially, the pressure is large (especially in the bubble) but decreases about an order of magnitude within the later evolutionary stages. While the results do not compare well with studies that are carried out in fewer dimensions \citep{Thornton1998, Blondin1998, Badjin2016} in terms of peak density in the Sedov-Taylor phase and in terms of resolving fluid instabilities that are certainly important in supernova-remnant modelling, we note that we compare well to other grid code results in three dimensions \citep[e.g.][]{Kim2015} who find a maximum value between 3.3 and 3.5 in their respective Sedov-Taylor phase. However, we note that we focus on low metalicity environments while \citep{Kim2015} is focusing on solar metallicity environments. This leads to differences in blast wave evolution which we discuss in section \ref{sec:past_ST}.  Moreover, we note that we observe a clear drop in the pressure between bubble and shell which indicates some contribution of the thermal instability in our highest resolution runs, although we see only a drop of $0.5$ dex while higher resolution state-of-the-art simulations of detailed blast wave evolution report $1$ dex \citep[e.g.][]{Badjin2016} and we do not properly resolve it at the presented resolutions throughout the paper. However, as we carry out these simulations to test which supernova-remnant properties can be modelled in current high resolution galaxy formation simulations, resolving the thermal instability in the remnant is not the main goal of the paper. Nevertheless, it would be very interesting to study the detailed effect of the thermal instability on momentum and hot phase generation in higher resolution simulations with our code in future studies. In this context, we have to consider the inability of capturing the analytic predicted value of the density and the inability of fully capturing the thermal instability as a limitation that has to be noted. However, despite this limitations which can be resolved at higher resolution or in lower dimensionality, the ability to model radial momentum input and hot phase at our target resolutions is worth noting as it enables the galaxy scale studies of outflows that are self-consistently driven by the the supernova-feedback without the limitations that common subgrid-models are tied to. 

To understand the details of the blast waves evolution we investigate the remnant at different mass resolutions. In Figure \ref{fig:all_dens_res} we show the remnant in the CNM during the ST-phase for all four mass resolutions that we investigated. We note that the shock is well resolved at the highest mass resolutions for the simulations CNM-L4 and CNM-L3. While the shock is still sufficiently resolved in the adiabatic regime in CNM-L2 it is completely unresolved for CNM-L1. This is mainly due to the fact that we have a lower amount of resolution elements in the same spatial region to actually resolve the shock properly in the adiabatic regime. Therefore, to capture the shock within the ST-phase a mass resolution of smaller than $10$ solar masses is needed for the MFM-solver. We mark runs that are resolved (green check), weakly resolved (black dot) and unresolved (red crosses) in Table \ref{tab:Table1}.
Further, we note differences in the cooling dominated regime. At the highest mass resolutions a thin, cooling dominated shell is forming following the end of the ST-phase. The density within this supernova shell is up to $10$ times higher than what we would expect to see in a pure adiabatic simulation of the shock front. However, with lower resolution this dense cooling shell becomes less and less resolved and remains unresolved at a mass resolution of $100$ M$_{\odot}$. We note that this becomes even worse for a different ambient configuration. If we consider the runs in the CMM the shock is less well resolved even at higher mass resolution. In the last evolution stage of the remnant the density decreases again and starts to merge with the ambient medium for all mass resolutions. We note that in these simulations we cannot capture the phase in which the SN-remnant merges with the ambient medium as the configuration of our ambient medium is chosen to be of uniform density and zero velocity. Consequently, our ambient medium is not turbulent which would be necessary to properly capture the merging phase of the shock with the ambient medium.
\begin{figure*}
        \includegraphics[scale=0.4]{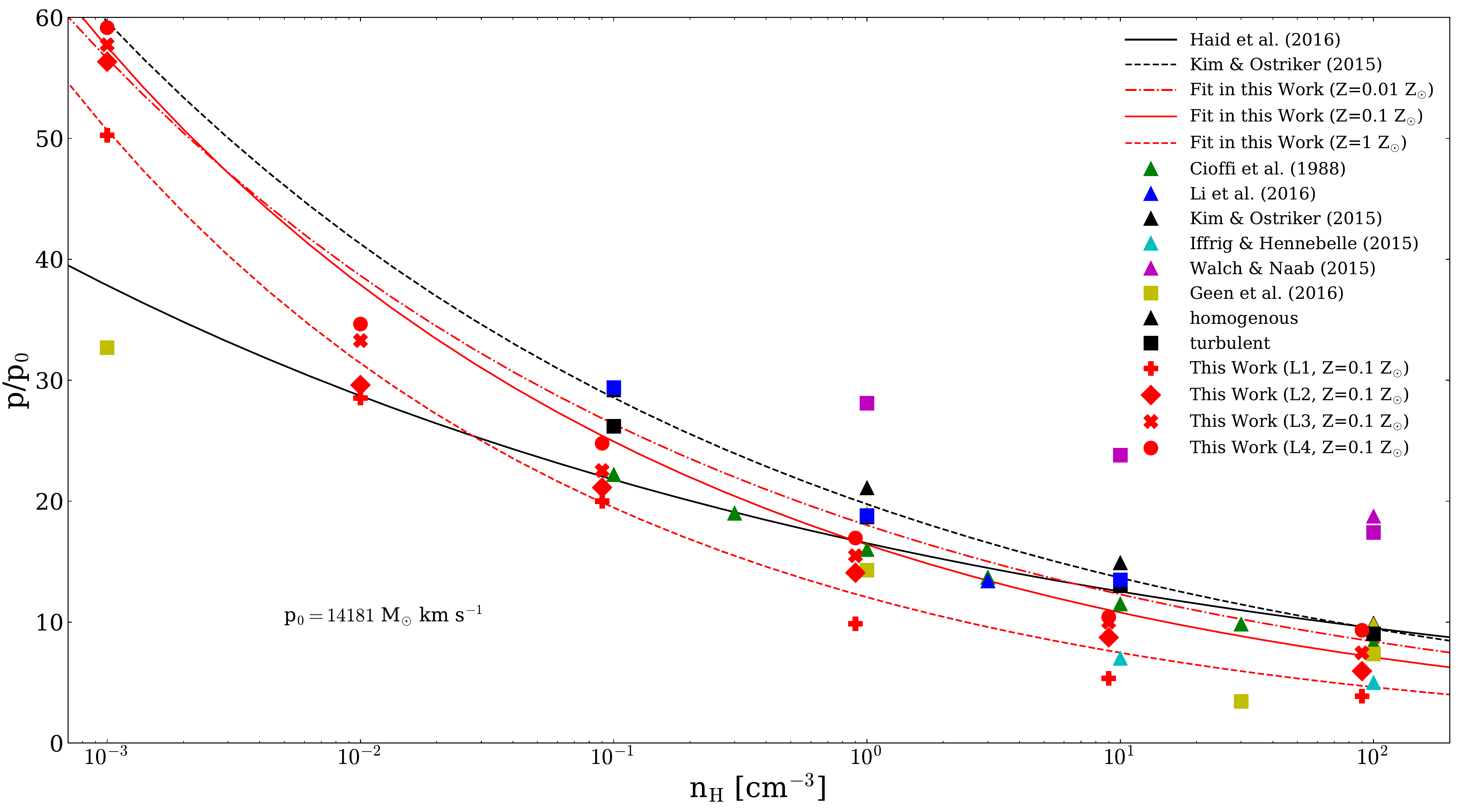}
        \caption{Momentum boost at the time of shell formation for blast waves at various environmental densities normalised to a fiducial initial momentum of p$_{0} = 14181$ M$_{\odot}$km s$^{-1}$. The coloured symbols show results from three dimensional numerical simulations (apart from \citet{Cioffi1988} and \citet{Haid2016}) with homogeneous (triangles), structured or turbulent (cubes) ambient media carried out with three different grid codes and a particle based SPH code. The red symbols show the supernova momenta obtained in this work at the end of the energy conserving ST-phase for our four different resolution levels in a homogeneous medium. At the lowest resolution the momenta can be underestimated by up to a factor 2. The lines show the fits from \citet{Haid2016}, \citet{Kim2015}, and the highest resolution of this work for all three adopted metallicities. We show the results obtained for a metallicity of $0.1$ $Z_{\odot}$ as red symbols for all environmental densities and mass resolutions. We note that we shifted the red symbols slightly to the left to show the results of other work below.}
        \label{fig:sn}
\end{figure*}

\subsection{Resolution effects}
\label{sec:resolution_effects}
For an accurate simulation of a supernova blast wave for applications in galaxy scale simulations it is important to capture the formation of the swept-up shell, which also determines the cooling properties of a supernova remnant \citep[e.g.][]{Kim2015}. For example \citet{Hu2016} have shown for SPH that the shell momentum is relatively insensitive to numerical resolution. However, the shell mass and velocity are strongly affected by higher velocities and lower shell masses at high resolution. At insufficient resolution the shell becomes too massive, slower and cools too early, resulting in a too rapid termination of the ST-phase. Similar results have been found by \citet{Kim2015} for Eulerian grid simulations. However, they focused on solar metallicity environments while we focus our study on sub-solar metallicities. The material in the shell has a temperature below $2 \cdot 10^{4}$ K.  All material above this temperature cut belongs to the hot phase of the remnant. Additionally, we set a velocity cut of 0.1 km s$^{-1}$ to select the ambient medium which is at rest form the bubble and the shell. The shell radius is defined by the position of the $10$ particles with the highest densities. The radial shell momentum is computed by summing up the individual particle momenta, and the shell velocity is the shell momentum divided by the shell mass. 

In Figure \ref{fig:1cmregion} we show shell radius (top left), shell momentum (top middle), shell energy (top right), split in kinetic (blue) and thermal energy (red), shell velocity (bottom left), shell mass and bubble mass (bottom middle), as well as the bubble temperature (bottom right). We present the four resolution levels with $0.1$ M$_{\odot}$ (solid line),  $1.0$ M$_{\odot}$ (dashed-line), $10$ M$_{\odot}$ (dash-dot-line),  $100$ M$_{\odot}$ (dash-dot-dot-line) for a medium with $n=1$ cm$^{-3}$ consisting of neutral hydrogen (CNM-L4 to CNM-L1). The small black dashes on the axis indicate the end of the ST-phase and the numerical values of the respective quantity for this point in time. We note that instead of showing the pressure in shell and bubble, we show the temperature structure within the hot bubble. The pressure within the bubble can be computed at any time by multiplying the densities within shell and bubble with the related temperatures. Moreover, we note that the lines for different resolutions start at different points in time because in the lower resolution runs the shock can not be separated from the ambient density field by the time at which the respective lines start. This is due to the fact that the shock remains unresolved until the respective points in time.
In the top left panel of Figure \ref{fig:1cmregion} before shell formation t < t$_\mathrm{sf}$ at the end of the Sedov phase, the radius is increasing as $r \propto t^{2/5}$ \citep{Sedov1946,Taylor1950} (green line). At shell formation (t$_\mathrm{sf}=45000$ yr) the slope changes $r \propto t^{2/7}$ for the momentum conserving phase. We define our Sedov-Taylor time t$_\mathrm{st}$ as the (final) time when the thermal energy drops below the analytical value of $72$ per cent. The three highest resolution levels capture the shell radius at formation and its evolution within $10$ per cent. The shell momentum (top middle panel of Figure \ref{fig:1cmregion}) is relatively insensitive to resolution and well captured at the three highest resolution levels. Further, our results are in good agreement with results from \citet{Kim2015} and \citet{Hu2016}. The shell velocity and mass, however, depend strongly on resolution (bottom left and middle panel of Figure \ref{fig:1cmregion}). With higher resolution the remnant velocity increases and the remnant mass decreases as the shock is better resolved. However, we note that a part of this trend is related to our feedback scheme. We further distinguish between the total mass of the remnant (black), the mass in the cold shell (blue) and the mass in the hot bubble (red). With decreasing resolution the mass within the bubble is lower apart from the lowest resolution run where the initially heated particles generate a too high hot mass. However, as the structure of the shell is so poorly resolved in these runs those particles cool very inefficiently and the hot mass is over predicted. For lower mass resolution we inject initially in more mass, leading to more initial swept-up material. This has consequences for the shell momentum at lower resolution. In the $10$ M$_{\odot}$ run the shell momentum is too high because of this reason, which renders the shell momentum a bad tracer to determine whether the ST-phase is resolved. The evolution of shell and bubble temperature is shown in the lower right panel. At the two highest resolution levels the bubble temperatures are resolved to within $10$ per cent. At the lowest mass resolution the temperature structure of the hot bubble is unresolved.  

\subsection{Blastwave evolution past the ST-phase}
\label{sec:past_ST}
While we focus on the evolution of the blastwaves within the ST-phase to investigate how momentum and hot mass are generated in this evolutionary state we briefly want to discuss the transition from the ST-phase to the PDS and finally the MCS phase. However, we note two major differences to other studies \citep[e.g.][]{Kim2015}. Our main focus is on sub-solar metallicity environments, as they can be found in present day dwarf-galaxies or at higher redshifts. Moreover, our cooling-function is non static and we excluded the effect of cosmic-ray ionisation to study SN-feedback in high density and very low temperature environments. Thus, we find three times higher terminal momenta compared to solar metallicity studies with higher equilibrium temperatures in the ambient ISM. First, metal-line cooling is less efficient in lower metallicity environments. This leads to an increased momentum boost past the ST-phase as the bubble cools less efficiently. Second, as we exclude cosmic ray ionisation th equilibrium temperature is lower and we can gain a little momentum boost after the PDS-phase as the shell is pushing into the ISM until the ambient ISM reaches pressure-equilibrium with the shell. Thus our momentum input gives an upper limit for the momentum boost which will generally be lower in practice as we will discuss in \ref{sec:galaxies}. In Figure \ref{fig:momentum_conserving} we show two examples for the CMM (blue) and the WNM (black) without (dashed) and with (solid) cosmic ray ionisation heating of the ISM at a rate of $3 \cdot 10^{-17}$ s$^{-1}$. We see a decrease of the final momentum by roughly 20 per cent if we switch on cosmic ray ionisation heating as the ambient pressure is higher and the shell cannot generate momentum anymore while pushing into the ambient ISM. Further, the exclusion of the cosmic-ray ionisation rate has an effect on the chemistry. While the molecular hydrogen is not effected (less than the per cent level), the cosmic-ray ionisation leads to higher fractions of ionised hydrogen by a roughly $10$ per cent and subsequently reduces the fraction neutral hydrogen. There is no effect on the formation of CO. Finally, we note that at the two highest mass resolutions we start to observe the fragmentation of the shell by Ryu-Vishniac type of instabilities. While we do not claim that these instabilities are fully resolved at the presented resolution, we note that they can further affect the cooling-rates that we calculate on the fly in our simulations. A detailed modelling of the cooling rate in shell and bubble plays a crucial role in modelling a supernova-event in a galactic context. The Ryu-Vishniac instabilities (even though our results are not fully converged yet) only appear in more than  one dimensional simulations and thus it is important to model the SN-feedback event in three spatial dimensions as they can influence the detailed cooling rates in our chemical model. We show this for our highest resolution simulation in a time evolution of four different snapshots, in the CMM-medium in density of 100 cm$^{3}$ in Figure \ref{fig:blastwave_evo}. We can see clear fragmentation of the remnant at very late times which is potentially caused by the Ryu-Vishniac instabilities. However, at this resolution the results could be also influenced by amplification of initial numerical noise of our particle distribution. We carefully checked the results by applying higher time resolution (i.e. decreasing the Courant number by a factor of 4) and find similar results. Moreover, we carried out a higher resolution simulation with a particle mass resolution of $10^{-4}$ M$_{\odot}$ and find similar structures.  
 
\subsection{Environmental dependence \label{sec:CMM}}
The evolution of SN remnants is strongly dependent on the environment. The cooling-time scales as $n^2$ which has consequences for remnants in a high density environment as well as remnants in a low density environment and leads to various different shell formation times. In Figure \ref{fig:sedov_time} we show the ambient densities for all our SN-remnants at the highest resolution (L4) as a function of the shell formation time t$_\mathrm{sf}$. With increasing ambient densities the shell formation times become shorter and can last from a few $1000$ years at $100$ cm$^{-3}$ to roughly $1.5$ Myr at $0.001$ cm$^{-3}$. As we simulate in three different metallicity regimes we show our best fits for the shell formation time individually for each metallicity as the dashed red line ($Z=1 Z_{\odot}$), the solid red line ($Z=0.1 Z_{\odot}$) and the dashed-dotted red line ($Z=0.01 Z_{\odot}$). For $Z=0.1 Z_{\odot}$ we additionally show the data from our six highest resolution runs to visualise the data with which we obtained the red solid line. The best fit relations as functions of the metallicity are given by:
\begin{align}
    t_\mathrm{st}(Z=0.01 Z_{\odot})  = 5.3 \cdot 10^{4} \mathrm{yr} \cdot n^{-0.50},  
    \label{eq:t_Z001}
\end{align}
\begin{align}
    t_\mathrm{st}(Z=0.1 Z_{\odot}) = 4.2 \cdot 10^{4} \mathrm{yr} \cdot n^{-0.50},
    \label{eq:t_Z01}
\end{align}
\begin{align}
    t_\mathrm{st}(Z=1 Z_{\odot}) = 2.9 \cdot 10^{4} \mathrm{yr} \cdot n^{-0.53}.
    \label{eq:t_Z1}
\end{align}
We compare these shell formation times to previous numerical and analytical results by \citet{Blondin1998}, \citet{Cioffi1988}, \citet{Franco1994}, \citet{Petruk2006} and \citet{Haid2016} and find good agreement with the recent results of \citet{Haid2016} and \citet{Petruk2006}. 

In Figure \ref{fig:various_densities} we show the blast wave evolution for all environmental densities at the highest resolution level (L4) corresponding to $0.1$ M$_{\odot}$ for different properties of the individual SN-remnants for an ambient metallicity of $Z=0.1 Z_{\odot}$. With increasing ambient number density the size of the remnants at the end of the ST-phase becomes shorter and can vary over more than an order of magnitude. The momentum at the end of the ST-phase is the largest for remnants in low density environments. For all environments apart from the highest ambient density we find a trend of an increasing kinetic energy in the PDS-phase of the remnant while the thermal energy is radiated away quickly after the end of the ST-phase. The shell velocities are the highest in the low density media where the shell and bubble masses are the highest. Further, we find that the bubble temperature is higher for higher mass densities after $10$ shell formation times, while the shell temperature significantly drops due to the shorter cooling times. Finally, we note that we can resolve the ST-phase for all ambient densities at our highest mass resolution. At lower mass resolution convergence can still be obtained for the three highest mass resolutions up to an ambient density of 10 cm$^{-3}$. For higher ambient media a mass resolution of $1$ M$_{\odot}$ is required to resolve the ST-phase. We show the results for a lower density environment and our highest ambient density environment in Appendix \ref{sec:appendixA}.

The shell formation time terminates the radial momentum generation phase (see \citealt{Haid2016} for a detailed discussion of the momentum generating phases) and the remnant transits to the momentum conserving snow plough phase. However, we note that given our cooling function and initial conditions our simulations indicate that the momentum is further increasing past pressure balance of bubble an shell is reached. However, as we have relatively low ambient temperatures due to our cooling function that does not contain the contribution of cosmic ray heating. While this provides relative extreme conditions with high cooling rates to determine how well the code can handle blast waves in rapidly cooling environments we lack comparability towards other studies like \citet{Kim2015} who employ a static cooling function with a higher equilibrium value for the temperature. For the high density environments we thus observe that the dense shell pushes further into the ISM and thus increases the momentum further until it reaches pressure equilibrium with the ambient ISM. In reality, this would be prevented by the much more complex structure of the ISM (e.g. overlapping remnants or interactions with stellar wind or magnetic fields), which we will discuss in more detail in section \ref{sec:galaxies}. For the remnants in the lower density media the situation is different as they have an equilibrium temperature of around $10^{4}$ K in the surrounding ISM, resulting in quickly after the ST-phase established equilibrium state between shell and ambient medium, where we then fully capture the momentum conserving snow-plough phase.

In Figure \ref{fig:sn} we show an updated version of Figure 4 of \citet{Naab2017} with a comparison of the momentum gain for different ambient densities to previous estimates by \citet{Cioffi1988, Martizzi2015, Iffrig2015, Li2017,Kim2015, Walch2015, Geen2016}. Our results agree well in particular with \citet{Kim2015} (despite a different cooling function) and show a momentum boost compared to the assumed initial SN momentum of p$_0$ = 14181 M$_{\odot}$ km s$^{-1}$ from a factor $\sim 10$ at $n = 100$ cm$^{-3}$ to a factor $\sim 50$ at $n = 0.001$ cm$^{-3}$. At low resolution this boost can be lower up to a factor of two at the highest densities. As we carried out all simulations at three different ambient metallicities, we can obtain best fit relations for all three metallicities that are given by: 
\begin{align}
    p_\mathrm{st}(Z=0.01) = 18.0 \cdot p_{0} \cdot n^{-0.16},
\end{align}
\begin{align}
    p_\mathrm{st}(Z=0.1) = 16.4 \cdot p_{0} \cdot n^{-0.18},
\end{align}
\begin{align}
    p_\mathrm{st}(Z=1) = 12.0 \cdot p_{0} \cdot n^{-0.17}.
\end{align}
These fitting relations are shown as the red dashed-dotted ($Z=0.01 Z_{\odot}$), the red solid ($Z=0.1 Z_{\odot}$) and the red dashed line ($Z=1 Z_{\odot}$) in Figure \ref{fig:sn}. The momentum is always measured at the end of the ST-phase. We note that a remnant can still generate momentum in the pressure driven snow-plough phase until pressure equilibrium between the bubble and the shell is reached, which can lead to higher momenta by a factor of two \citep{Haid2016, Kim2015}. Compared to other studies \citep[e.g.][]{Haid2016, Kim2015} we follow the momentum gain for $Z=0.1 Z_{\odot}$ and $Z=0.01 Z_{\odot}$. We find a weak dependence of the final terminal momenta generated by each remnant in different metallicity environments which mostly affects the terminal momenta in the high ambient density regimes (roughly a factor of 2) while the momenta in the lowest ambient environments remain unchanged.    
\begin{figure*}
        \includegraphics[scale=0.5]{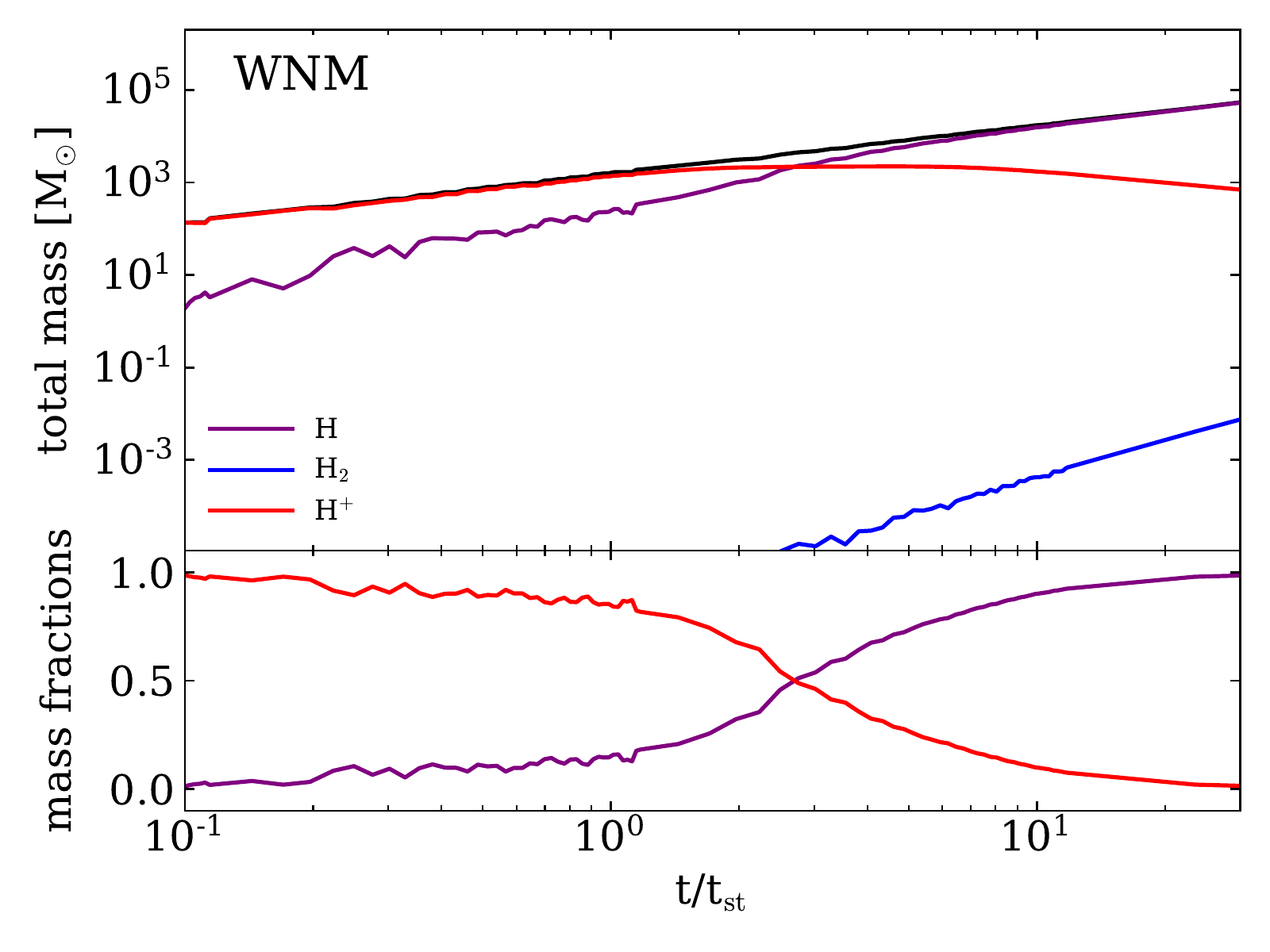}
        \includegraphics[scale=0.5]{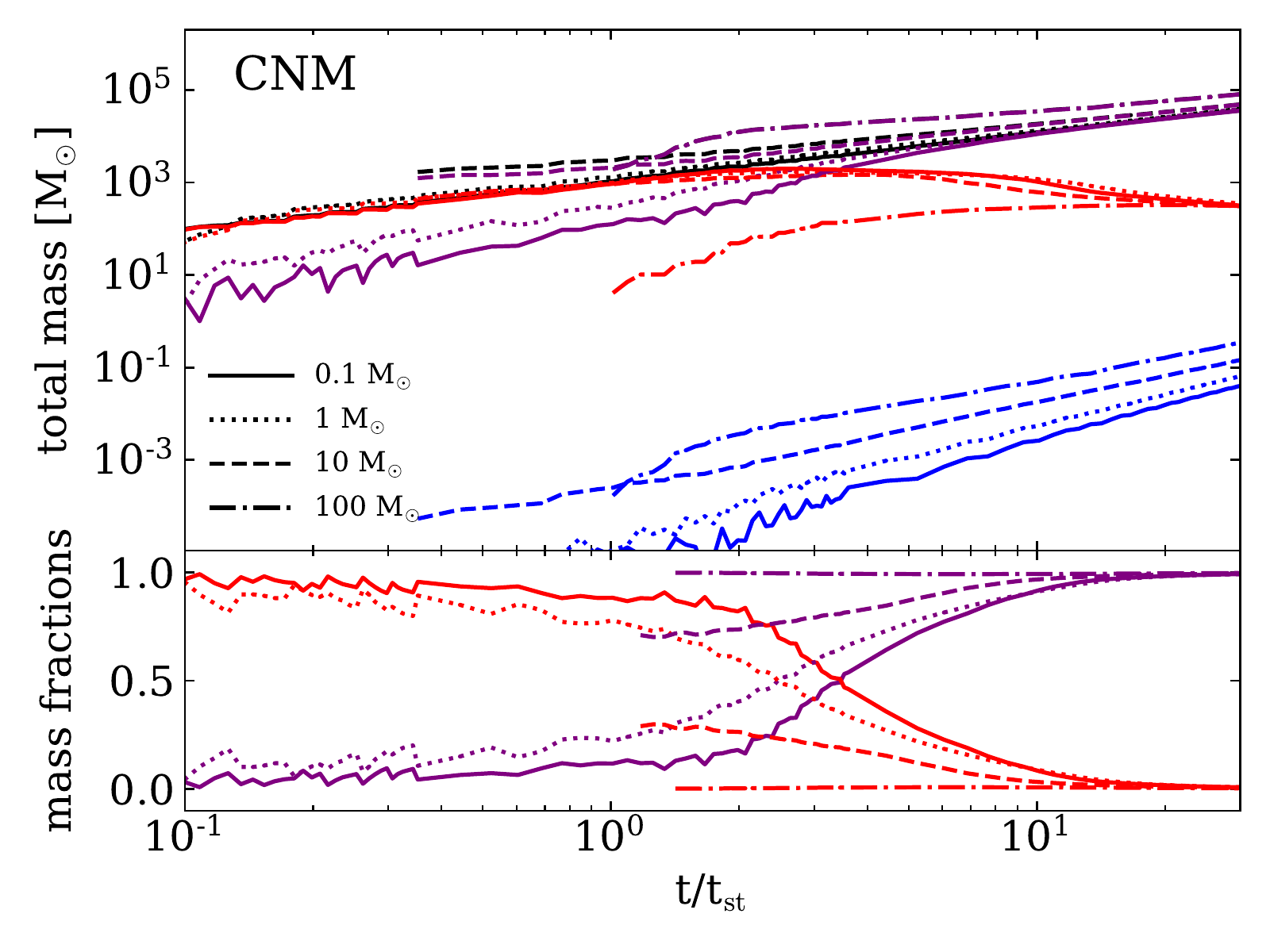}
        \includegraphics[scale=0.5]{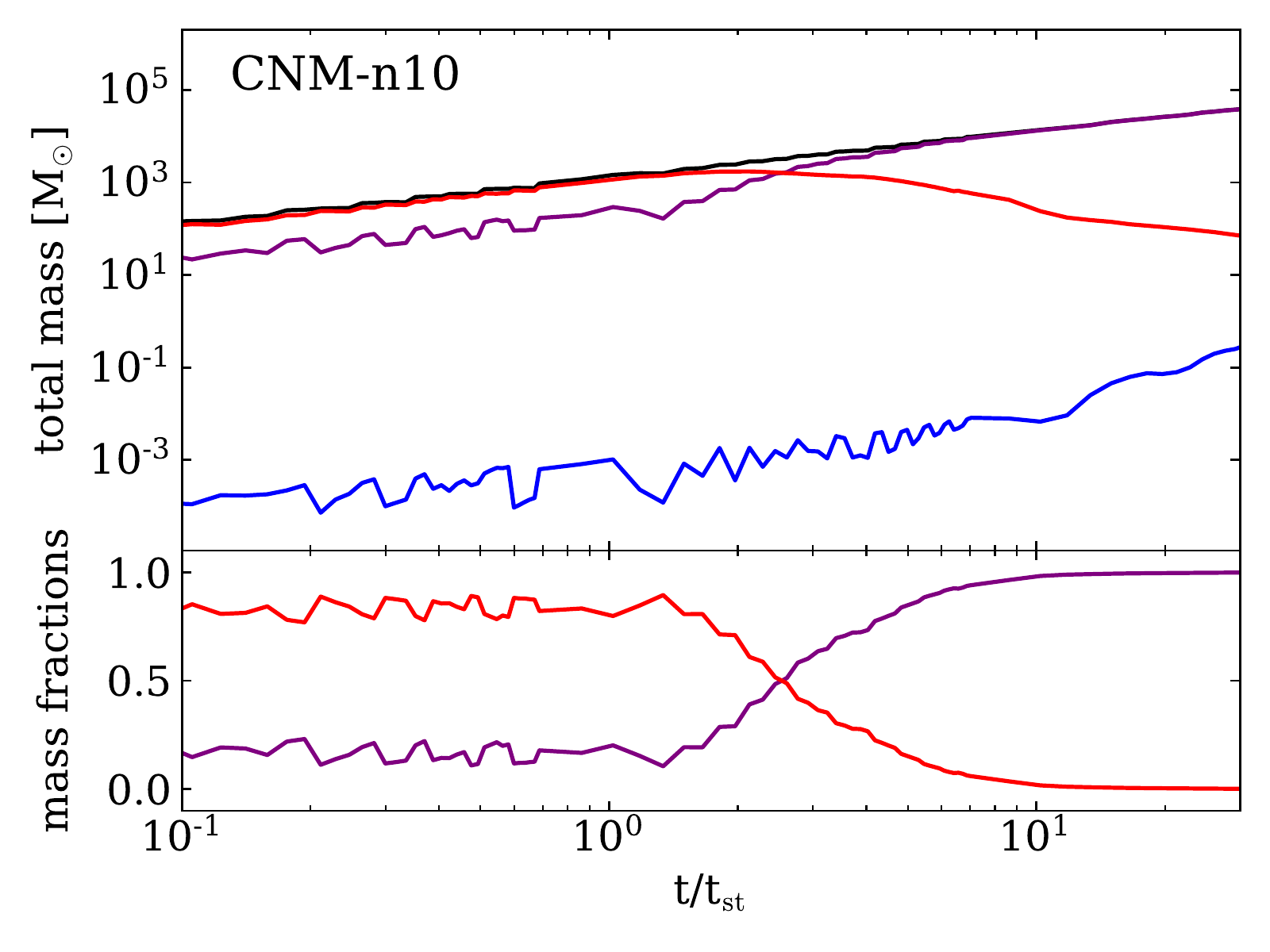}
        \includegraphics[scale=0.5]{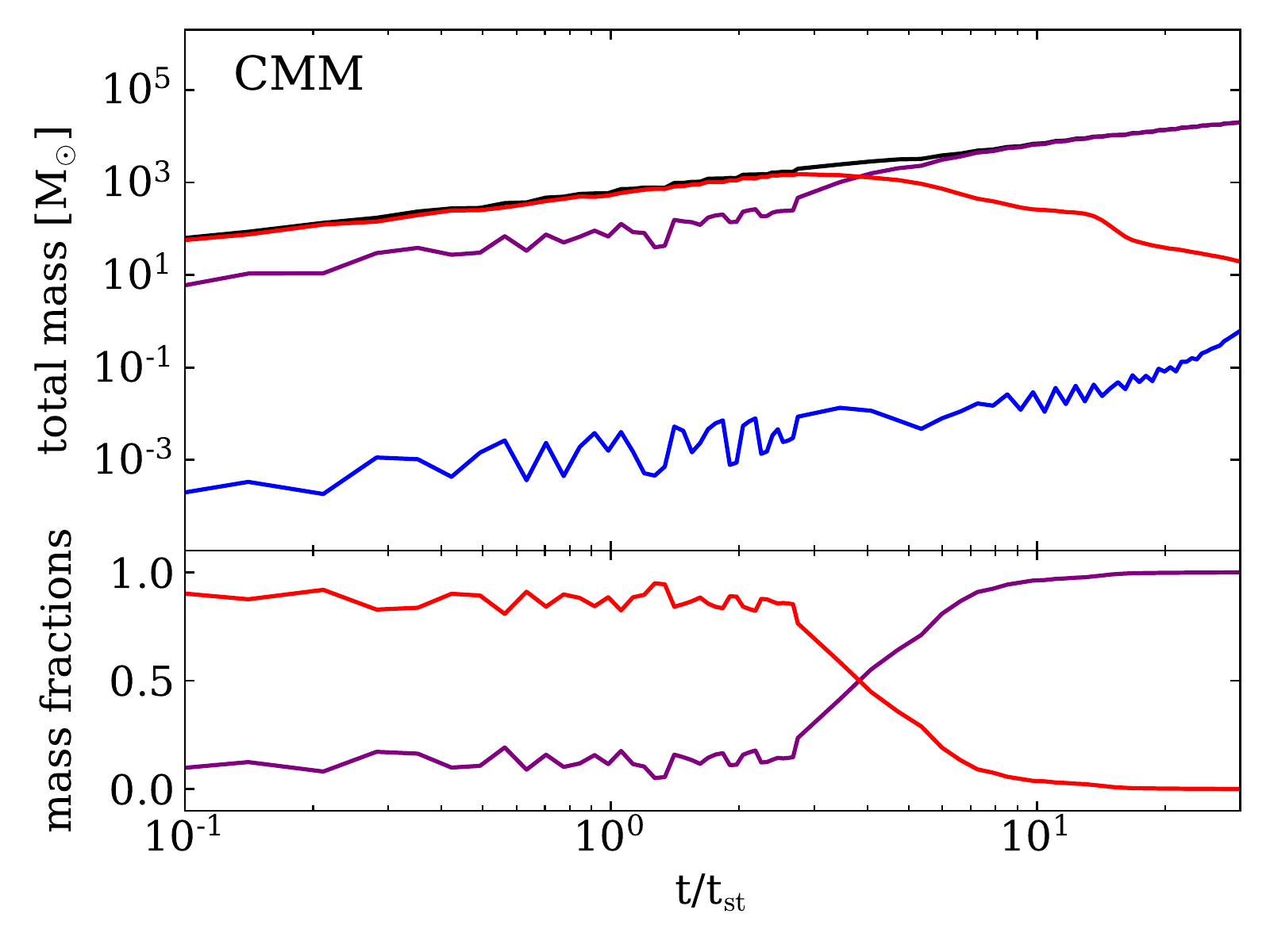}
        \caption{Chemical composition of the masses of SN-remnants in different environments of the non-equilibrium species H$_{2}$ (blue), H$^{+}$ (red) and H (purple) for the runs WNM-L4 (top right), CNM-L4 (top right), CNM-n10-L4 (bottom left) and CMM-L4 (bottom right) as a function of time. The top panels show the total mass in each species while the bottom panels show the mass fraction of each species. We observe that initially most of the molecular and neutral hydrogen is destroyed due to the heating of the injection energy. The remnants in the high density environments can preserve a small fraction of their mass in neutral hydrogen. At later times the shell becomes fully neutral again in all remnants. In all cases we can see that a little mass in molecular hydrogen builds up. However, it is below the per cent level within $10$ Sedov-times. The black lines indicate the total mass in the remnant.}
        \label{fig:abundances}
\end{figure*}
\subsection{Chemical composition}

\begin{figure}
        \includegraphics[scale=0.5]{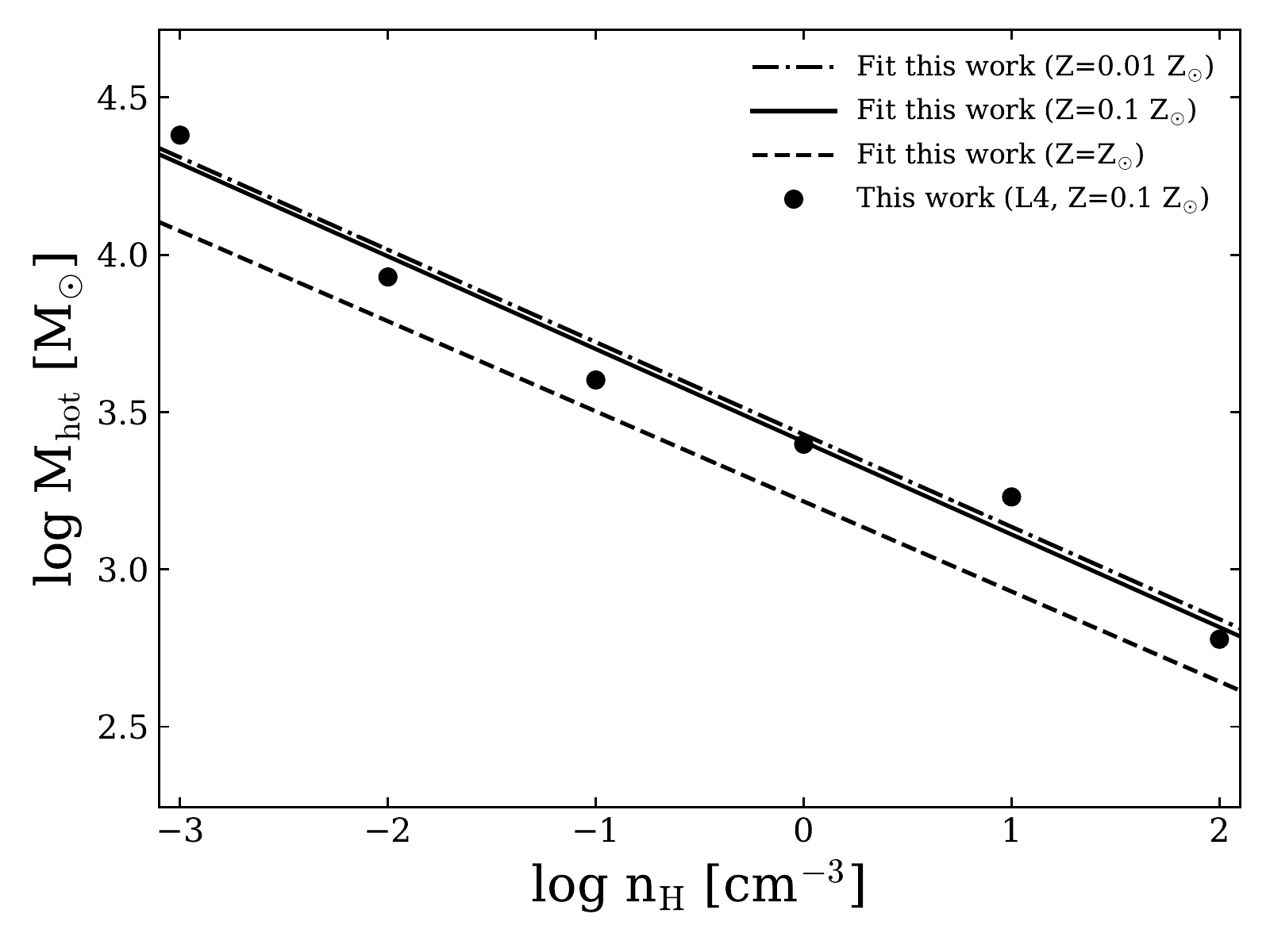}
        \caption{Peak hot mass as a function of the environmental density in which the SNe occur. We show the peak hot mass for a metallicity of Z=0.1 Z$_{\odot}$ as red dots. The red dashed, the red solid and the red dash-dotted line are the best fit relations that are obtained for the metallicities Z=Z$_{\odot}$, Z=0.1 Z$_{\odot}$ and Z=0.01 Z$_{\odot}$. We find that remnants in low density generate one order of magnitude more hot gas mass as remnants in high density environments.}
        \label{fig:hot_mass}
\end{figure}

We show the chemical evolution of SN-remnants for H$_{2}$, H, and H$^{+}$ in the WNM, the CNM, the CNM-n10, and the CMM in Figure \ref{fig:abundances}. In the WIM (top left) the initial material is fully ionised and H$^{+}$ stays the dominating chemical component of the shell even after $2$ shell formation times where roughly $60$ per cent of the remnant are in full ionisation. However, at ten shell formation times the remnant becomes fully neutral and also forms a tiny fraction of molecular hydrogen (below a solar mass). The CNM (top right panel) initially consists of neutral hydrogen but the whole remnant is ionised by the blast wave. Only after shell formation the remnant cools and the shell is becoming 100 per cent neutral again after $10$ shell formation times. In these remnants molecular species can be ignored. For the CNM, we show a resolution study for the chemical fractions within the remnant. At the lowest mass resolutions the chemical composition of the single remnant is not captured accurately. At this resolution the remnant is not fully ionised in the beginning and initially more mass is kept in the neutral state rather than fully ionised by the blast wave as expected. The remnant is fully neutral before shell formation. The same is true for the $10$ M$_{\odot}$ runs, although we note that here we already find a slightly dominating ionisation fraction of the remnant compared to the neutral component. For the two highest resolution runs we find that the blast wave is fully ionising the ambient medium, keeping 70 to 80 per cent of the gas ionised until shell formation. After that, cooling is dominating and the remnant starts to form neutral hydrogen and reaches the fully neutral state roughly $10$ shell formation times after the appearance of the blast wave. Despite the fact that the molecular fractions are low, they also vary within at least an order of magnitude at the end of the simulation, where the low resolution remnants produce more molecular hydrogen than the high resolution ones, which leads to a more dominant cooling process through molecules at low resolution. This indicates that the non-equilibrium chemical model overestimates the cooling rate at low resolution increasing the issues with the well known cooling problem.

In higher density media (CNM-n10) we find the same behaviour (bottom left panel of Figure \ref{fig:abundances}). However, we note that initially a small fraction of hydrogen of around 20 per cent is kept in the neutral phase and the formation rate of molecular hydrogen is slightly enhanced (half an order of magnitude) towards the end of the simulation.  
In the CMM medium (bottom left of Figure \ref{fig:abundances}) we find that the some of the initially swept up mass is kept in the neutral state (again roughly 20 per cent) while the rest if fully ionised. The formation rate of molecular hydrogen increases again and at the end of the simulation we find about half an order of magnitude lower molecular fractions than ionised rest mass of the remnant. If we continued the simulation longer the built-up of molecular hydrogen would be significant. However, we assume that remnants overlap before they reach more than $30$ Sedov-times.\\
Finally, we briefly discuss the dynamical impact of the chemical model. We carry out one reference run where we override the evolution of the chemical abundances and only enable our non-equilibrium cooling routines. Not evolving the chemical abundances has mainly an impact on the cooling behaviour of the gas. We only find a marginal impact of the chemical model on the overall evolution of the supernova remnant on the percent level. However, the chemistry effects the initial momentum build-up and the peak hot mass in the beginning. The initial momentum is slightly increased in the run without the active chemical evolution, but converges quickly to the momentum evolution that is observed with the active chemistry. The peak hot mass is slightly increased without the chemical evolution, which can be explained by the missing cooling channel over the molecules. Furthermore, we want to point out that the radiation which is emitted by the remnant would certainly impact its chemical evolution which is a potential caveat of the model. However, the simulations are also too low in resolution to make valuable prediction for X-ray properties of supernova-remnants. To accurately investigate this, one would have to carry out higher resolution simulations with lower dimensionality like \citet{Thornton1998}, \citet{Blondin1998}, or more recent \citet{Badjin2016}. However, We investigate a possible feedback scheme that aims for the input of the terminal momentum and the generation of the hot phase in galaxy scale simulations in a resolved fashion. Although the chemistry might be insignificant in isolated events that picture can be quickly altered if we consider multiple events that lead to overlapping higher density remnants with shorter cooling times and higher formation rates for molecular hydrogen, which will certainly influence the distribution of internal energy and therefore its dynamics within the ISM.

\subsection{Comparison of different Hydro-solvers \label{sec:mfv}}

\begin{figure*}
        \includegraphics[scale=0.5]{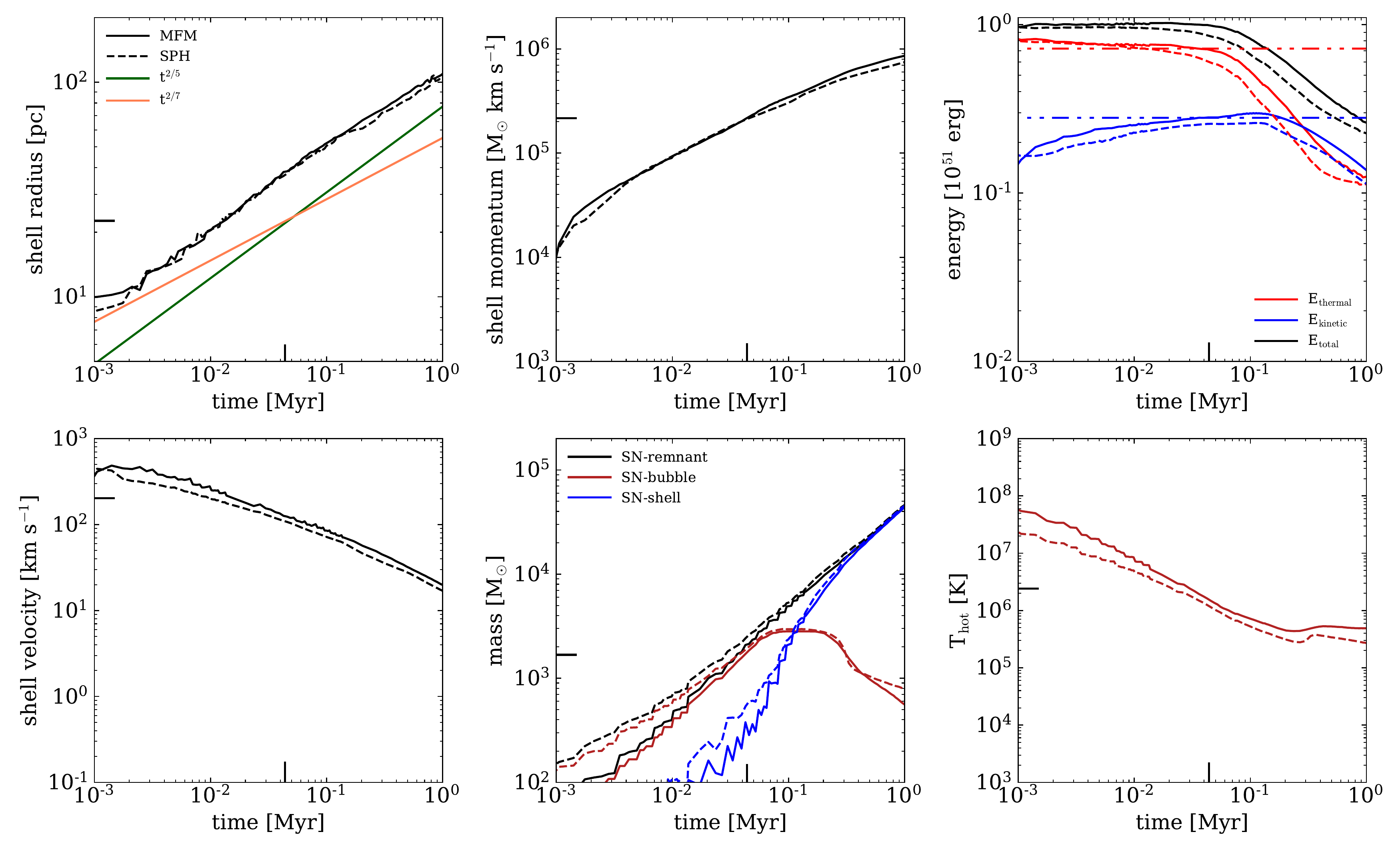}
        \caption{Comparison of the shock capturing behaviour for two different solvers, MFM (solid) and SPH (dashed). Most physical properties agree within a few percent.}
        \label{fig:comp_solvers}
\end{figure*}

We carried out some of our simulations employing different solvers for hydrodynamics. All simulations at our highest mass resolution are carried out with our pressure-energy SPH solver. In Figure \ref{fig:comp_solvers} we show the comparison between the MFM-solver (solid) and the SPH-solver (dashed) for all physical quantities that we investigate. Overall, we find good agreement of the two solvers at the highest mass resolution. Most of the physical properties of the shock are captured within a few per cent. The major differences between the solvers are given in the behaviour of the thermal and kinetic energy (upper left panel of Figure \ref{fig:comp_solvers}). For the SPH-solver we find that the thermal energy is radiated away at a slightly earlier time compared to the MFM solver leading to shorter shell formation times (c.f. equation \ref{fig:sedov_time_cond}). Further, the SPH-solver under predicts the kinetic energy within the ST-phase. The other difference that we point out is that the SPH-solver predicts a lower temperature in the hot bubble by half an order of magnitude. Therefore, the pressure within the bubble is slightly lower, which can potentially lead to less momentum generation within the PDS phase for this method. For the SPH-solvers we evaluate the results in section \ref{sec:iso_cond} in more detail.

\section{Build up of the hot phase}
\label{sec:semi}

Following \citet{Naab2017} we can derive the expectation value for a hot phase to be established by SN-feedback. To derive the expectation value we follow a semi-analytic approach. From our simulations we can derive the radii of the remnant as a function of the environmental density. As we do not have data for different explosion energies we lack the dependence on the explosion energy compared to similar studies \citep[e.g.][]{Kim2015}. However, we tested different metallicities and can therefore carry out the semi-analytic calculations for different metallicity regimes. First, we need to derive the dependence of the radius on the ambient density. We can fit our results for all three different regimes. Therefore, we obtain the radius of the shell at the end of the ST-phase. We find the following relations.
\begin{align}
  r_\mathrm{st}(Z=0.01 Z_{\odot}) =  18.0 \ \mathrm{pc}\ n^{-0.53}  
  \label{eq:r_Z001}
\end{align}
\begin{align}
  r_\mathrm{st}(Z=0.1 Z_{\odot}) =  16.4 \ \mathrm{pc}\ n^{-0.52}
  \label{eq:r_Z01}
\end{align}
\begin{align}
  r_\mathrm{st}(Z=1 Z_{\odot}) =  13.15 \ \mathrm{pc}\ n^{-0.52}
  \label{eq:r_Z1}
\end{align}
The expectation value for the hot phase of a SN-explosion that goes off within the hot phase of a previously occurred SNe can be written as follows
\begin{align}
    \epsilon_\mathrm{hot}=S \frac{4 \pi}{3} r_{st}^{3}t_{st},
\end{align}
where $S$ is the SN-rate and r$_\mathrm{st}$ is the remnants radius at the end of the ST-phase.  By substituting r$_\mathrm{st}$ with the equations \ref{eq:r_Z001} to \ref{eq:r_Z1} and t$_\mathrm{st}$ with equations \ref{eq:t_Z001} to \ref{eq:t_Z1} we find a strong power law dependence for the expectation value of the hot phase for all three different metallicity regimes. The expectation value for the hot phase can then be determined via:
\begin{align}
    \epsilon_\mathrm{hot}(Z=0.01 Z_{\odot}) = S \cdot 1.3 10^{-6} \mathrm{kpc}^3 \mathrm{Myr}^{-1} n^{-2.1},
\end{align}
\begin{align}
    \epsilon_\mathrm{hot}(Z=0.1 Z_{\odot}) = S \cdot 9.74 10^{-7} \mathrm{kpc}^3 \mathrm{Myr}^{-1} n^{-2.1},
\end{align}
\begin{align}
    \epsilon_\mathrm{hot}(Z=1 Z_{\odot}) = S \cdot 6.62 10^{-7} \mathrm{kpc}^3 \mathrm{Myr}^{-1} n^{-2.1}.
\end{align}
Thus, we see that the expectation value for the hot phase varies only very weakly with the metallicity and we can determine it for typical values within the ISM. We assume a typical surface density of the Milky Way within the solar neighbourhood around 10 M$_{\odot}$ kpc$^{-2}$. We follow \citet{Naab2017} and assume a Salpeter mass function and a disc height of 250 pc. From that we can determine the SN-rate $S$ for solar neighbourhood conditions and obtain 280 kpc$^{-3}$ Myr$^{-1}$. By assuming the typical medium density within the Milky Way of $n=1$cm$^{-3}$ this gives a very low number for the expectation value for the hot phase in order of $10^{-5}$ to $10^{-4}$. However, for SNe in lower density environments the expectation value for the hot phase becomes quickly larger and for environmental densities below $0.01$ we find an expectation value greater than one due to the strong power law dependence of the expectation value with the density of the ambient medium. We note that this has consequences for the hot phase in the ISM. Once a configuration of the ISM is reached where SNe go off in low density environments our model predicts higher and higher expectation values with decreasing densities. SNe tend to go off in lower and lower density media. Cooling times for the gas increase alongside with the sizes of the remnants. Finally, this becomes a runaway process with dominating volume filling hot phase. This picture is in very good agreement with the findings from our simulations. Further, the model is in good agreement with the simulations carried out by \citet{Girichidis2016} who investigated the gas densities in galactic outflows and find a peak of the number density of roughly $0.01$ cm$^{-3}$. While the remnants in the CMM have a size of a few pc, the size of remnants in the warm and hot ionised phase of the ISM can reach values up to $800$ pc. 
Moreover, in Figure \ref{fig:hot_mass} we show that the hot mass of the remnant significantly increases as a function of the environment. We measured the peak hot mass (maximum mass of the bubble) that is generated by SN-blastwaves in different environments and find strongly increasing masses in the hot phase in remnants below $0.1$ cm$^{-3}$. We find the following best fit relations. 
\begin{align}
    M_\mathrm{hot}(Z=0.01 Z_{\odot}) = 2570 \cdot M_{\odot} \cdot n^{-0.29},
\end{align}
\begin{align}
    M_\mathrm{hot}(Z=0.1 Z_{\odot}) = 2398 \cdot M_{\odot} \cdot  n^{-0.29},
\end{align}
\begin{align}
    M_\mathrm{hot}(Z=1 Z_{\odot}) = 1584 \cdot M_{\odot} \cdot  n^{-0.28},
\end{align}
Therefore, remnants in low density environments contribute much more to the build up of the hot phase by  generating high expectation values for the hot phase alongside with high mass fractions of the hot gas. Remnants in high density media on the other hand have very low expectation values for a hot phase to form and generate only around $600$ M$_{\odot}$ of hot gas in the bubble which quickly cools away once the shell forms.

This has consequences for the formation of a wind driven by SNe. In this picture a wind is driven by the hot phase via the formation of a superbubble which interior is heated by SNe expanding into the bubbles of preceding SNe. Once the bubble breaks out of the disc the pressure within the bubble interior can push the shell outwards into the CGM. Hereby the outflow velocity would be limited by the sound speed within the CGM. Given a virial temperature of $10^6$ K for the CGM of the Milky Way this can drive outflows with a few 100 km s$^{-1}$ as they are observed \citep[e.g.][]{Genzel2011}.     

\section{Effects of thermal conduction}
\label{sec:iso_cond}

Using our fiducial SPH-solver we investigate the behaviour of SN-remnants under the effect of thermal conduction.  
In supernova remnants, thermal conduction can influence the interface between the hot bubble and the shell by redistributing thermal energy from the bubble into the shell and mass from the shell into the bubble \citep{Keller2014,El-Badry2019}. With a temperature dependent thermal conduction coefficient in the Spitzer-limit (see Sec. \ref{sec:thermal_conduction}) we investigate differences in remnant morphology and chemistry. All our highest resolution runs have been carried out including the effect of thermal conduction.   
\begin{figure*}
        \centering
        \includegraphics[scale=0.5]{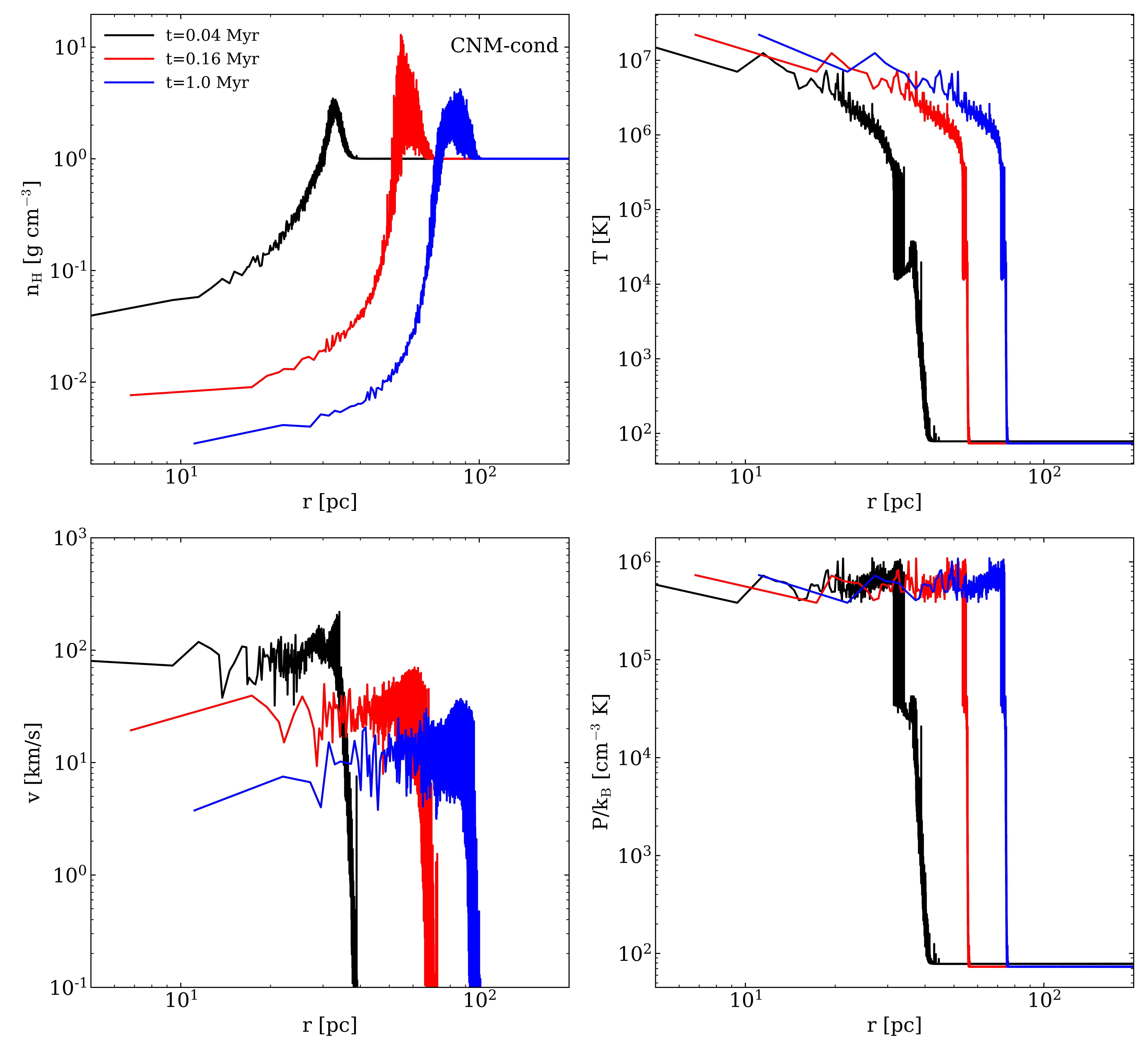}
        \caption{Same as Figure \ref{fig:1cmregion} but with the effect of isotropic thermal conduction. We find slightly lower temperatures in the bubble and higher densities in the shell.}
        \label{fig:1cmregion_cond}
\end{figure*}
In Figure \ref{fig:1cmregion_cond} we show the radial profiles of the density (top left), the temperature (top left), the velocity (bottom right) and the pressure (bottom right) for the simulation CNM-Lv4. Comparing to Figure \ref{fig:all_dens}, we find that the temperature in the run with thermal conduction is about an order of magnitude lower than in the case without thermal conduction, while the pressure remains roughly constant. The lower bubble temperatures result form heat flux from the hot bubble to the colder shell. The pressure is similar due to mass flux from the shell to the bubble. 
This effect has a general impact on the evolution of the remnants when thermal conduction is included. 

\begin{figure}
        \includegraphics[scale=0.5]{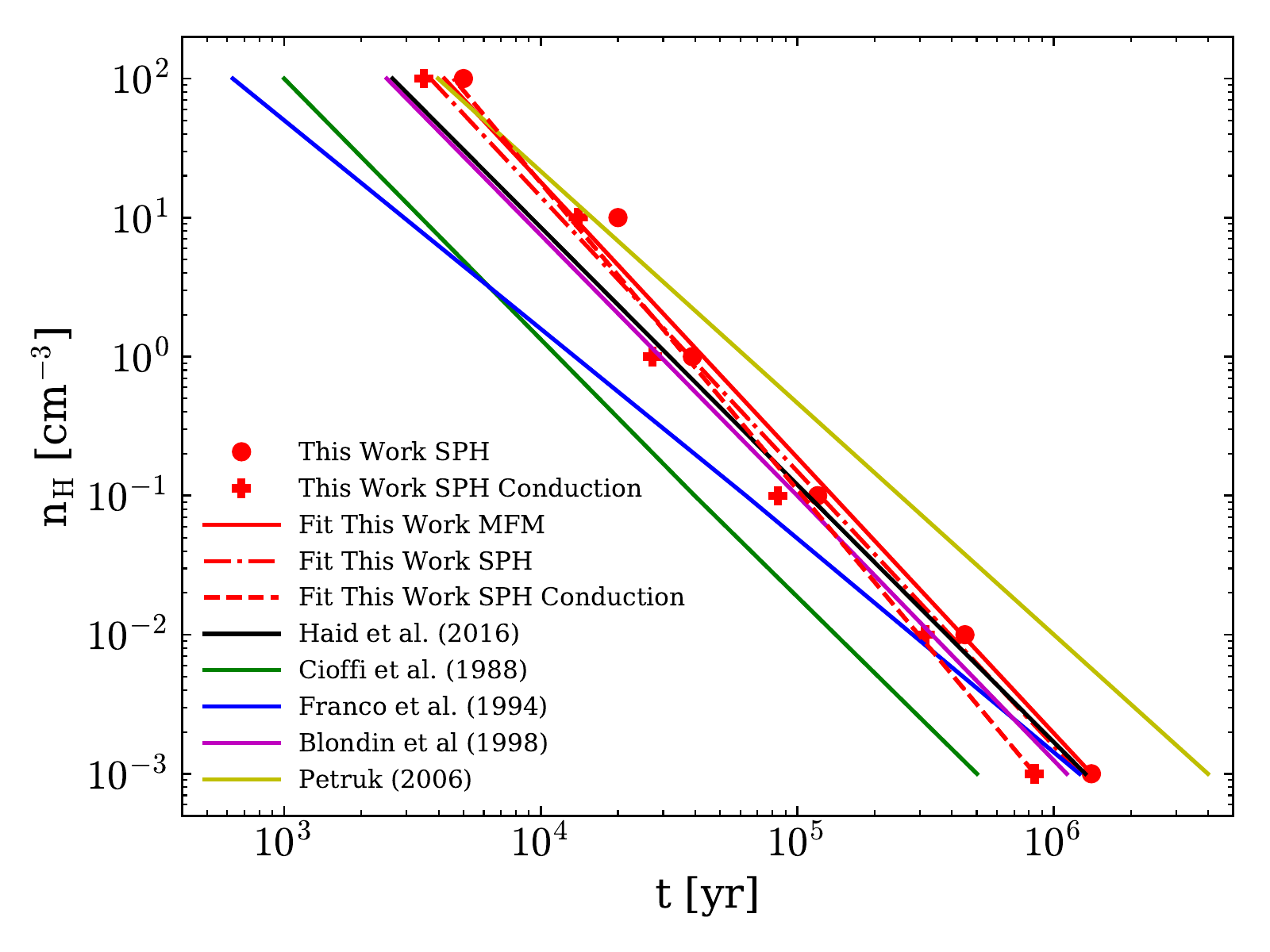}
        \caption{Same as Figure \ref{fig:sedov_time} for the results that we obtain with SPH (red dashed) and SPH with heat conduction (red dashed dotted). For reference we overplot the results we obtained with MFM. We find slightly shorter shell formation times for the SPH-solver than we find for the MFM solver. Moreover, thermal conduction further shortens the shell formation time.}
        \label{fig:sedov_time_cond}
\end{figure}

\begin{figure*}
        \includegraphics[scale=0.4]{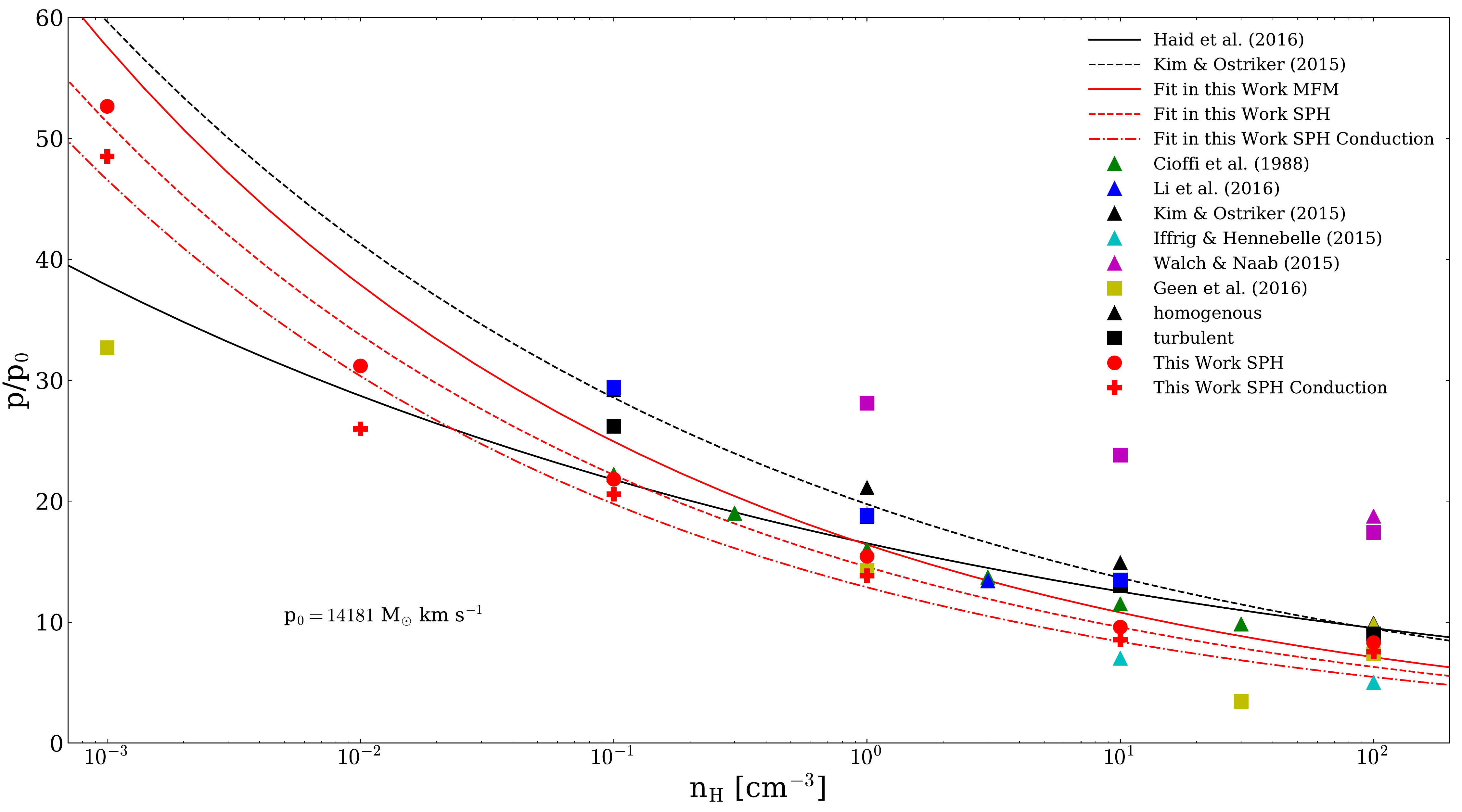}
        \caption{Same as Figure \ref{fig:sn} for the runs with SPH (red dashed) and thermal conduction (red dashed dotted). We directly overplot the data points we obtained for SPH (red dots) and for the runs including heat conduction (red crosses). In comparison with MFM we generally find that the momentum is reduced by roughly $10$ per cent for SPH and $20$ per cent with SPH and heat conduction.}
        \label{fig:size_cond}
\end{figure*}

\begin{figure}
        \includegraphics[scale=0.5]{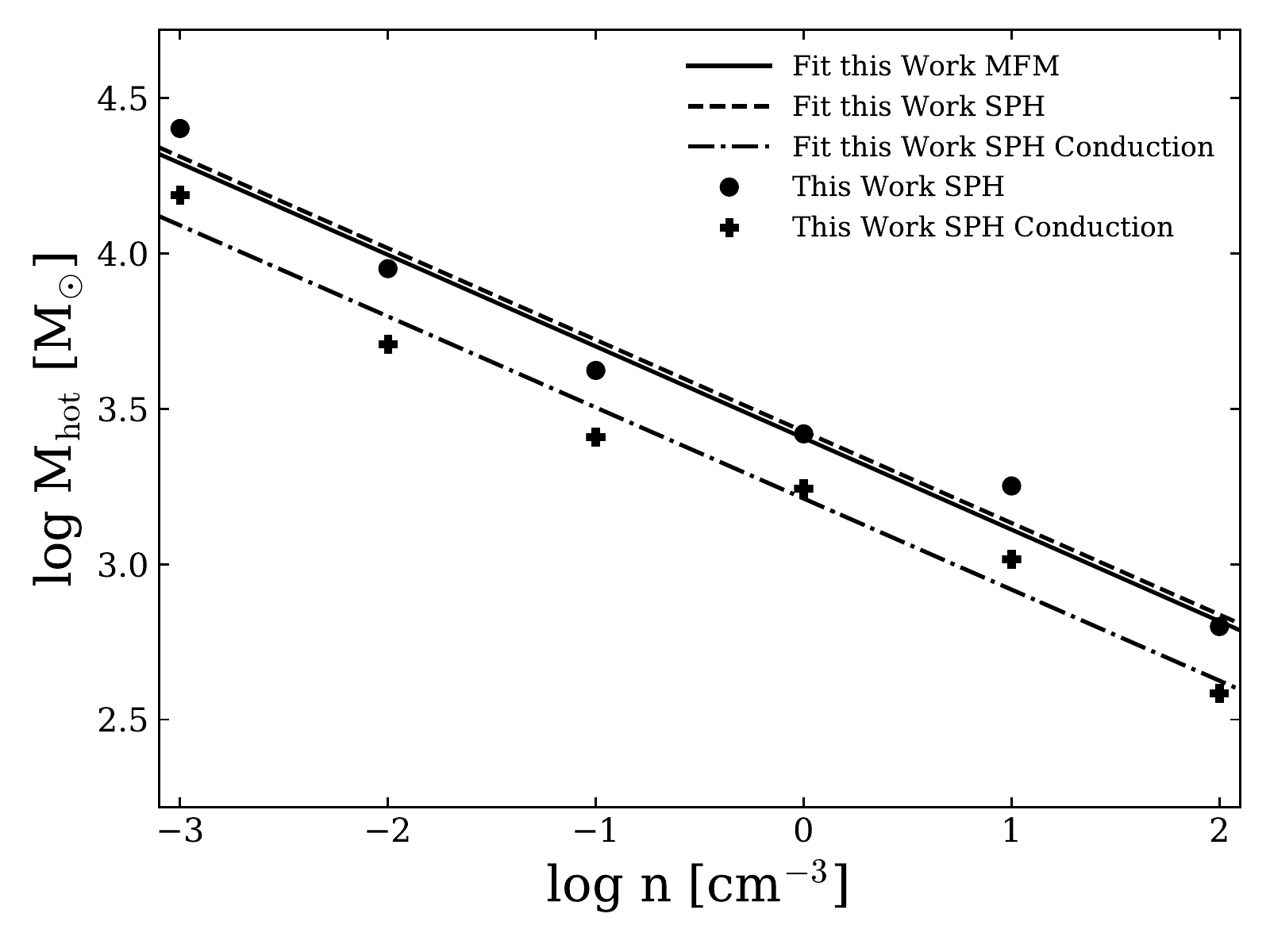}
        \caption{Same as Figure \ref{fig:hot_mass} for the runs that are carried out with SPH (dashed) and with SPH and conduction (dashed-dotted). While we only find weak differences between MFM and SPH in the peak hot mass we find that thermal conduction can reduce the peak hot mass by roughly $40$ per cent.}
        \label{fig:hot_mass_cond}
\end{figure}
Conduction describes a heat flux from the hot gas to the cold gas, which is counterbalanced by a mass flux from the cold gas to the hot gas constrained by energy conservation. This is valid before shell formation and leads to slightly higher mass fractions of the hot gas. However, in SN-remnants cooling is active after shell formation and energy conservation is violated. Therefore, the mass flux from cold to warm gas is prohibited and the energy from the bubble that enters the shell is radiated away immediately. This can impact the general physical properties of SN-remnants. Temperature is reduced and density slightly increases.    
While there are quantities that are only weakly affected by thermal conduction (e.g. remnants size, velocity structure), there are others which are affected significantly. First, we note that the ST-times (times of shell formation) are shorter in the presence of thermal conduction by roughly $20$ per cent. We show this in Figure \ref{fig:sedov_time_cond} where we plot a comparison of the shell formation times that we find in our fiducial runs with MFM (red solid line), the SPH (red dashed line) and with SPH plus thermal heat conduction (red dahsed dotted line). Conduction leads to a redistribution of internal energy from the hot gas to the cold. This reduces temperatures and leads to a slightly enhanced cooling. Therefore, cooling times shorten and the energy conserving ST-phase terminates earlier. We find modified fitting formulas with conduction. For completeness we give the fit we obtained with SPH.
\begin{align}
    t_\mathrm{st, sph}  = 3.8 \cdot 10^{4} \mathrm{yr} \cdot n^{-0.50},  
    \label{eq:sph}
\end{align}
\begin{align}
    t_\mathrm{st, cond} = 3.6 \cdot 10^{4} \mathrm{yr} \cdot n^{-0.46}.
    \label{eq:sph_cond}
\end{align}

Because of the earlier termination of the ST-phase, the momentum input with thermal conduction is lowered compared to the runs without thermal conduction. In Figure \ref{fig:size_cond} we show the momentum input as a function of the environmental density and find that the SPH solver gives lower momenta by roughly $10$ per cent while thermal conduction again reduces the terminal momentum by $20$ per cent compared to the results that we obtain with our fiducial MFM solver. We overplot the data points from the simulation as red dots (SPH) and red crosses (SPH plus thermal conduction). The modified fitting formulas for the results with SPH and conduction yield:
\begin{align}
    p_\mathrm{st, sph} = 14.6 \cdot p_{0} \cdot n^{-0.18},
\end{align}
\begin{align}
    p_\mathrm{st, cond} = 12.9 \cdot p_{0} \cdot n^{-0.18}.
\end{align}
However, the largest impact can be seen in the build up of the hot mass which is reduced by 40 per cent. We show the results in \ref{fig:hot_mass_cond} for MFM (solid line), SPH (dashed line) and conduction (dashed-dotted line). While the results for MFM and SPH only differ in the per cent regime, the hot mass is reduced $40$ per cent in the runs with thermal conduction. We note that before cooling takes over the hot mass slightly increased due to the mass flux from warm to hot gas. For thermal conduction we find the dependence of the hot mass as function of the environmental density as follows:
\begin{align}
    M_\mathrm{hot, cond} = 1580 \cdot M_{\odot} \cdot n^{-0.29}.
\end{align}
\begin{figure}
        \includegraphics[scale=0.5]{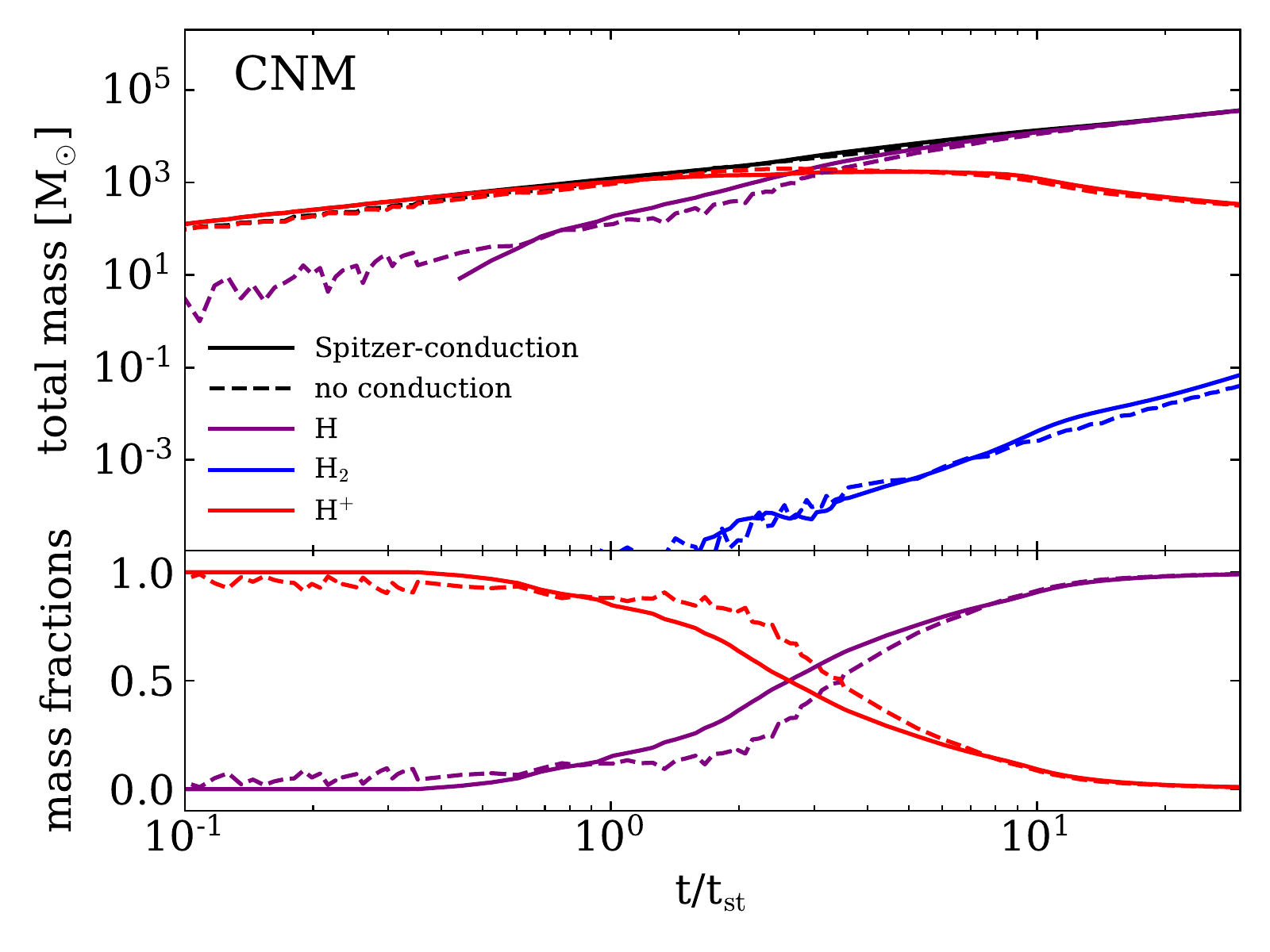}
        \caption{Comparison between the chemical evolution of the shell without the effect of thermal conduction (solid) and with thermal conduction (dashed) for the different species that the non-equilibrium model follows. We find slightly increased fractions for the H$^{+}$ in the runs that include thermal conduction in the beginning of the simulation. At later times the ionising fraction of hydrogen decreases faster than in the run without thermal heat conduction.}
        \label{fig:cond_chem}
\end{figure}

Further, we find an impact of thermal conduction on the chemical composition of the SN-remnant. Overall, the effect is minor. Therefore, we show the results for the run CNM-cond in Figure \ref{fig:cond_chem}. Initially, the majority of the gas is ionised in the case with and without conduction but with conduction the ionising fraction is lightly reduced (by roughly 5 per cent). Because cooling is slightly more efficient in the presence of conduction, the build-up of neutral hydrogen in the shell is slightly enhanced. This then leads to higher formation rates of molecular hydrogen which is increased by a factor of two. 

\section{Applications to astrophysical systems}
\label{sec:galaxies}

In this section we briefly discuss applications to larger scale ISM and galactic scale systems to show that it is possible to use the presented feedback scheme at the target resolution of solar mass and sub-parsec resolution and achieve convergence on the momentum and hot phase generation in a global large scale simulations just by resolving the momentum and hot phase of isolated remnants. Moreover, we provide an insight into SN-remnants in structured media which can not be handled in higher resolution one dimensional SN-blast waves studies as these systems miss the radial symmetry needed for justifying a lower dimensional approach.

\subsection{The supernova driven ISM}

\begin{figure*}
        \includegraphics[scale=0.9]{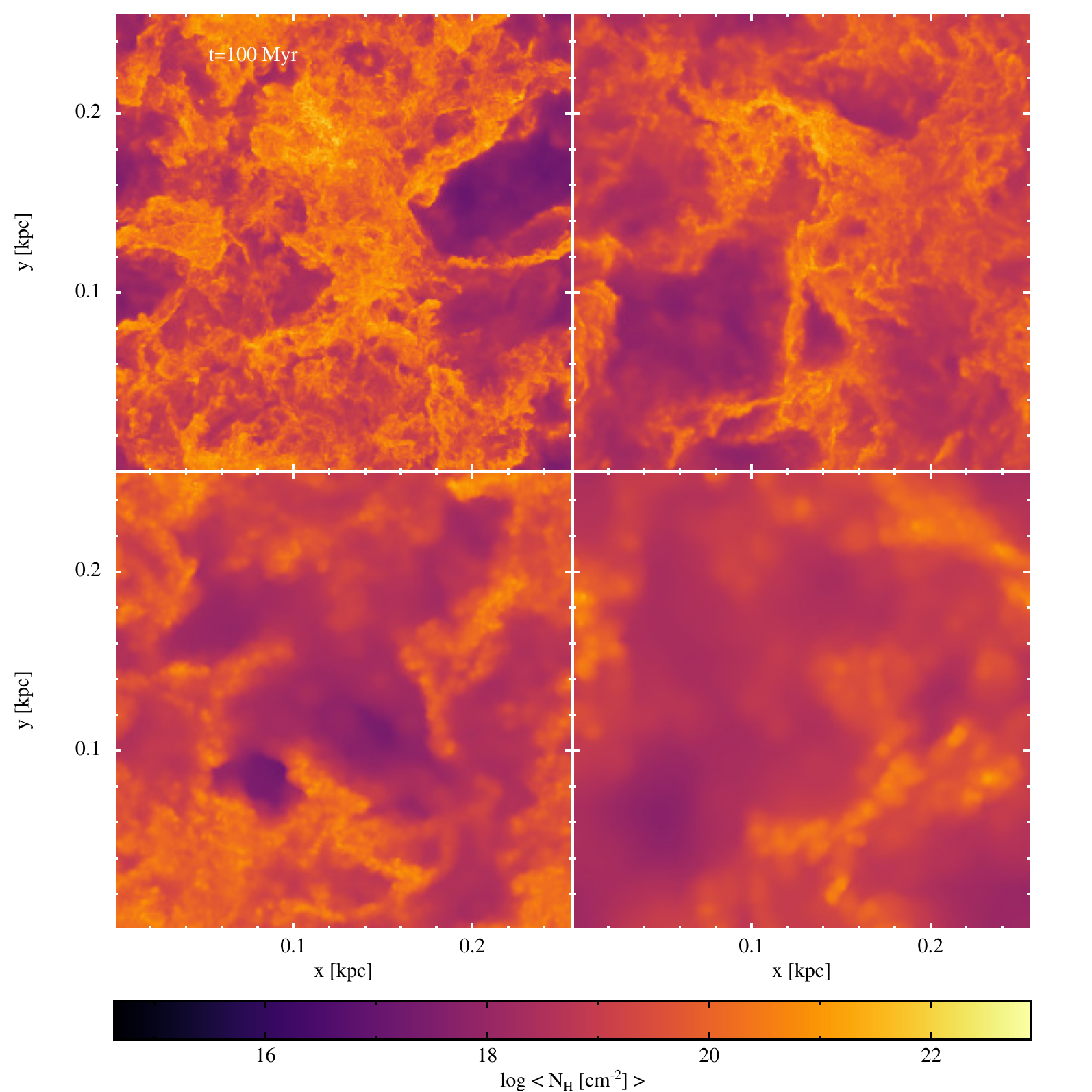}
        \caption{We show the effect of multiple supernova events on the resulting ISM-structure as a function of particle mass. The upper left panel shows our highest mass resolution of $0.1$ M$_{\odot}$, the upper right the $1$ M$_{\odot}$, the bottom left the $10$ M$_{\odot}$ and the bottom right the 100 M$_{\odot}$ resolution. While the two highest resolution runs show significant substructure from filaments and cloud like structures, the lower resolution runs with $10$ solar masses but specifically the $100$ solar mass run, show a far less structured and much more blurred out medium.}
        \label{fig:ISM-sn-driven}
\end{figure*}

\begin{figure}
    \centering
    \includegraphics[scale=0.5]{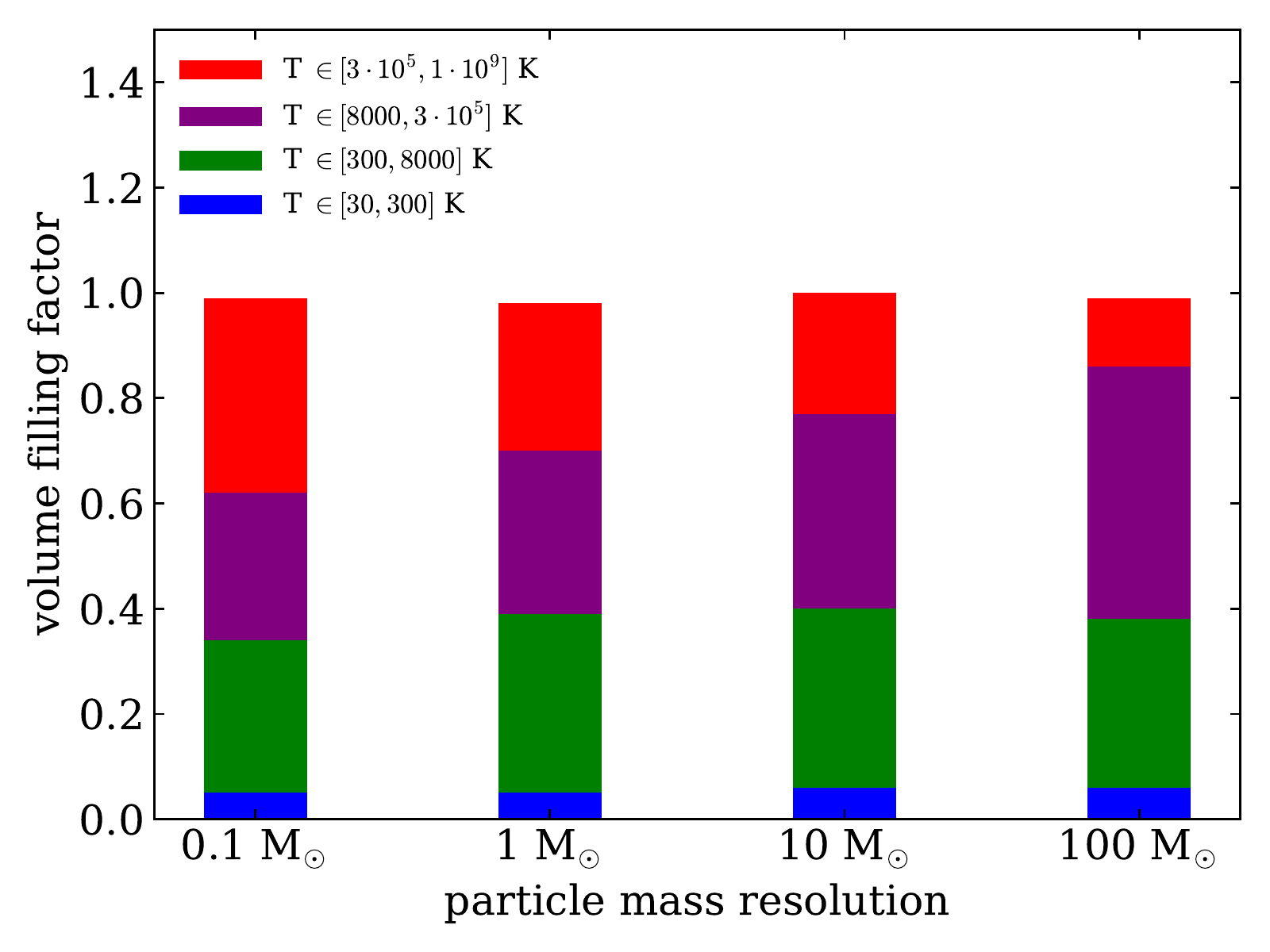}
    \caption{We show the mass weighted Volume filling fractions for the simulations that are shown in \ref{fig:ISM-sn-driven} for all different mass resolutions. For the highest mass resolutions it is possible to generate a Volume filling hot and warm phase as it is observed within the local ISM.}
    \label{fig:vff}
\end{figure}

\begin{figure*}
        \includegraphics[scale=0.8]{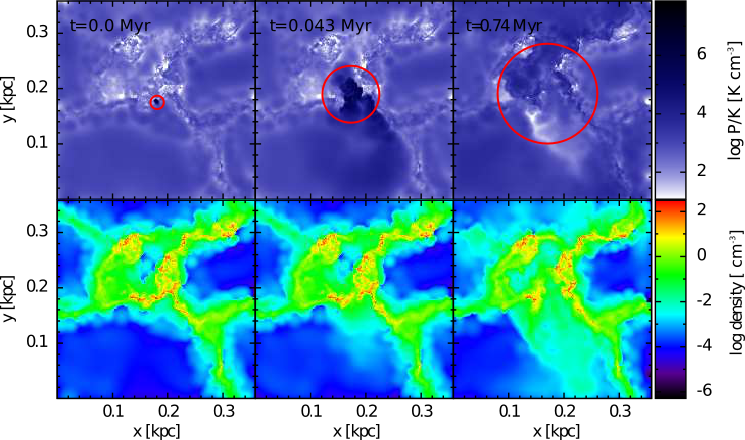}
        \caption{We show the evolution of a supernova-blast wave that evolves into a supernova-driven ISM as pressure slices (top row) and density slices (bottom row) at the height at which the blast wave event occurs (box center, $z=0.168$ pc). We show the irregular evolution of this blast wave for three different points in time. In the upper left panel we show the initial density configuration. The supernova-event is placed in the very centre of the box in the densest filament we could identify. The blast wave is then evolving into the direction of the steepest pressure gradient (in positive and negative y-direction). Alongside the filament the action of the feedback-event is suppressed due to the higher environmental densities in this direction. The middle panel shows the end of the ST-phase. We immediately see that the remnant looks much more irregular and shows much more detailed structure compared to the isolated events, capturing the full three dimensional nature of this problem. Past the ST-phase we observe the evolution of an irregular blast wave structure that is supported by the internal pressure provided by the hot material in the bubble (right panels), until the bubble is in pressure equilibrium with the shell at which point the remnant starts to merge with the ambient medium.}
        \label{fig:sedov_structured}
\end{figure*}

\begin{figure*}
        \includegraphics[scale=0.6]{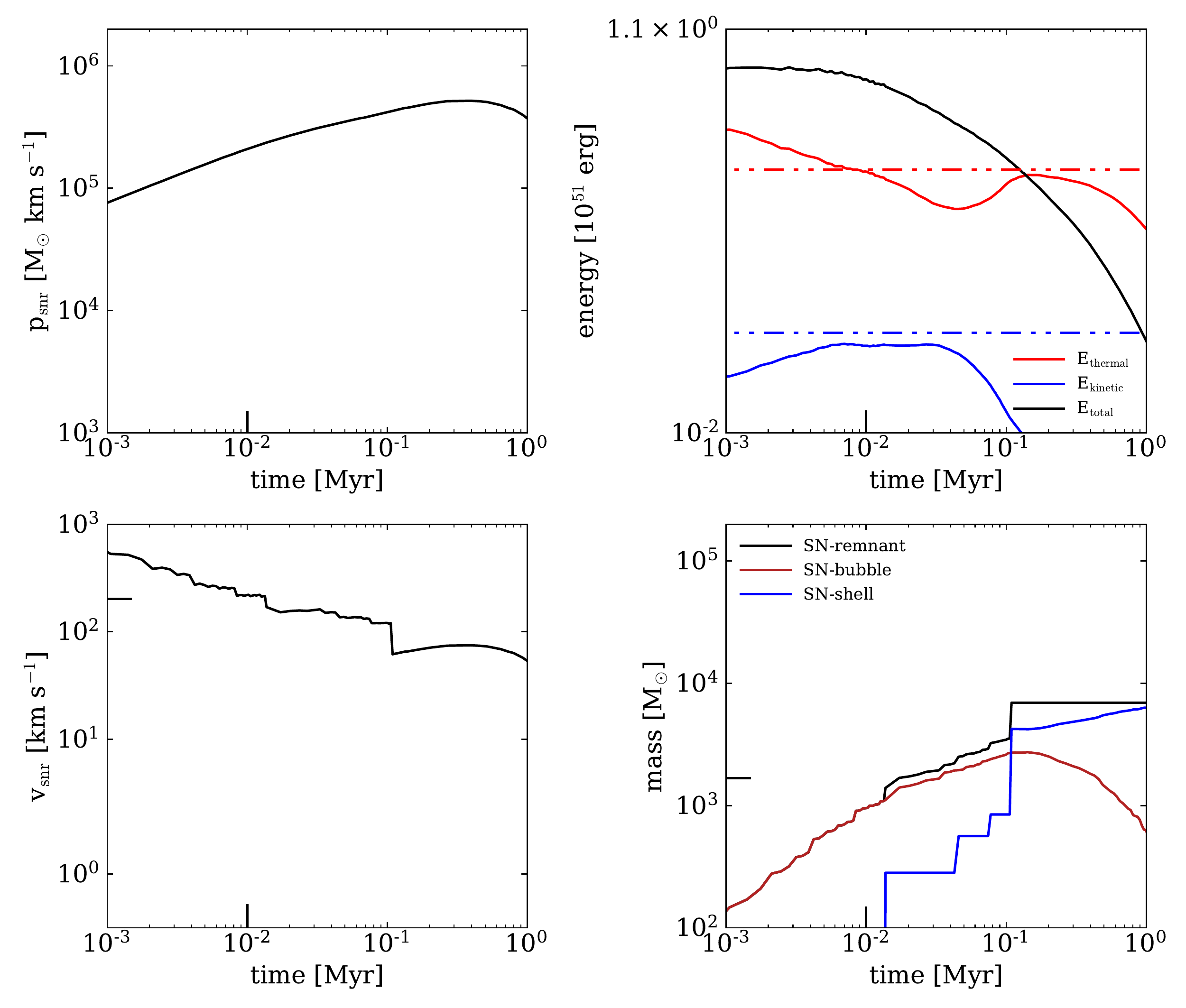}
        \caption{We show the most important SN-remnant properties (momentum, energy, velocity and mass) that can be modelled with a simple SN-feedback scheme at our highest resolution level. Generally we find that the remnant undergoes a proper ST-phase with a split of 72 \% in thermal and 28 \% in kinetic energy, before the cooling losses start to dominate and the remnant goes onto the pressure driven snow-plough phase before entering the momentum conserving phase which is indicated by a flattening of the momentum curve (bottom left). After this the momentum drops and starts to merge with the ambient (turbulent) ISM. Compared to the isolated events we observe roughly the same terminal momentum build-up towards the end of the ST-phase. The momentum is increased by a factor of two in the pressure-driven snow plough phase. The hot mass seems to be increased by a factor of three compared to the isolated events. This is related to the asymmetry of the remnant. In the structure medium feedback is most effective alongside the direction of the highest pressure gradient. Thus energy is deposited into lower density gas which can be heated very effectively, resulting in a more effective heating mechanism.}
        \label{fig:structured_event}
\end{figure*}

The first example we want to study is the SN-driven ISM. In Figure \ref{fig:ISM-sn-driven} we show the density distribution of the SN-driven ISM in a periodic box with a side length of 256 pc with solar neighbourhood conditions for the background UV-field but $0.1$ Z$_{\odot}$ which represents an ISM environment as present at high redshift. However, we adopt a lower SN-rate of 1.5 supernova per Myr and follow the evolution of the SN-driven ISM for 100 Myr. The four different panels in Figure \ref{fig:ISM-sn-driven} represent the four different mass resolutions after 100 Myr of evolution. Visually, we note some differences in density structure. In the lowest resolution run we observe that the resulting ISM-structure is only very badly resolved and the hot phase is mostly missing in this simulation as the canonical SN-energy falls to over cooling and is radiated away shortly after deposition without depositing momentum or generating the hot phase ans subsequently it does not provide the ISM-pressure to subsequently generate a SN-driven outflow.  Finally, we investigate the SN-remnant properties that we obtain from a feedback event in a structured (SN-driven) medium. In Figure \ref{fig:sedov_structured} we show the blast wave evolution for 4 different point of a blast wave that evolves into the SN-driven ISM for our highest resolution runs. In this scenario we generate the initial conditions by using the SN-driven ISM simulations that we show in Figure \ref{fig:ISM-sn-driven} and select a snapshot with a nicely build-up turbulent structure that provides an realistic environment for the explosion of massive star as we would obtain it in galactic scale simulation. The canonical SN-energy of $10^{51}$ is hereby distributed over the $32$ particles closest to the centre of the box. Thus the SN-even occurs in the centre of the box at $x=180$ pc and $y=180$ pc. The supernova goes of in the dense filamentary structure that can be observed in the upper left panel of Figure \ref{fig:sedov_structured}. With this event we represent a supernova-explosion in the CMM. As the blast wave occurs in the centre of a filament the feedback is most effective into the direction of the steepest pressure gradient. In this scenario this coincides with the steepest density-gradient. Therefore, we obtain a highly asymmetric blast wave structure with a faster expansion of the blast wave in negative and positive z-direction. While the blastwave can expand freely in negative y-direction it expands into an already bubble-like in positive y-direction and heats the material in this already present bubble even more (upper right panel of Figure \ref{fig:sedov_structured}). Despite the fact that the feedback is ineffective parallel to the filament we can still observe in the upper right panel of Figure \ref{fig:sedov_structured} that it get disrupted, by the momentum that is deposited during the ST-phase which remains resolved in this simulation. After the end of the ST-phase (bottom panels of Figure \ref{fig:sedov_structured}) the material in the irregular formed shell is further pushed out via the pressure-support by the hot material inside the newly formed bubble, generated in the resolved ST-phase.Finally, we show the detailed blast wave properties in Figure \ref{fig:structured_event} and point out the differences to the results of the runs that we obtained with a non-structured, constant density ambient medium at rest. The main difference is that in the context of an already structured ambient medium the heating by the supernova-explosion is most effective in the direction of the steepest pressure gradient. Thus the remnant becomes highly non-isotropic. While it is relatively easy to follow the momentum, the energy distribution and the hot mass fraction, of this non-isotropic remnant it is not straight forward to determine the cold mass as most of the other parts of the ISM are still cooling while they are not yet effected by the heating from the supernova. The velocity can only be determined as averaged value out of shell mass and shell momentum. We show the central quantities of the remnant in a structured medium in Figure \ref{fig:structured_event} with the momentum (top left), the energy (top right), the velocity (bottom left) and the remnant mass (bottom right). We can see that the remnant, despite its irregular shaped form still undergoes the ST-phase, the PDS-phase and the MCS-phase, which can be identified in the momentum and energy distribution. Velocity and cold mass are harder to track in this context while the hot mass can be determined more easily. we note that we find a factor of 2.5 more hot mass in the structured medium compared to the remnants in homogeneous media. This can again be explained by the more efficient heating alongside the steepest density gradient.

\subsection{Galaxy scale simulations}

\begin{figure*}
        \includegraphics[scale=0.3]{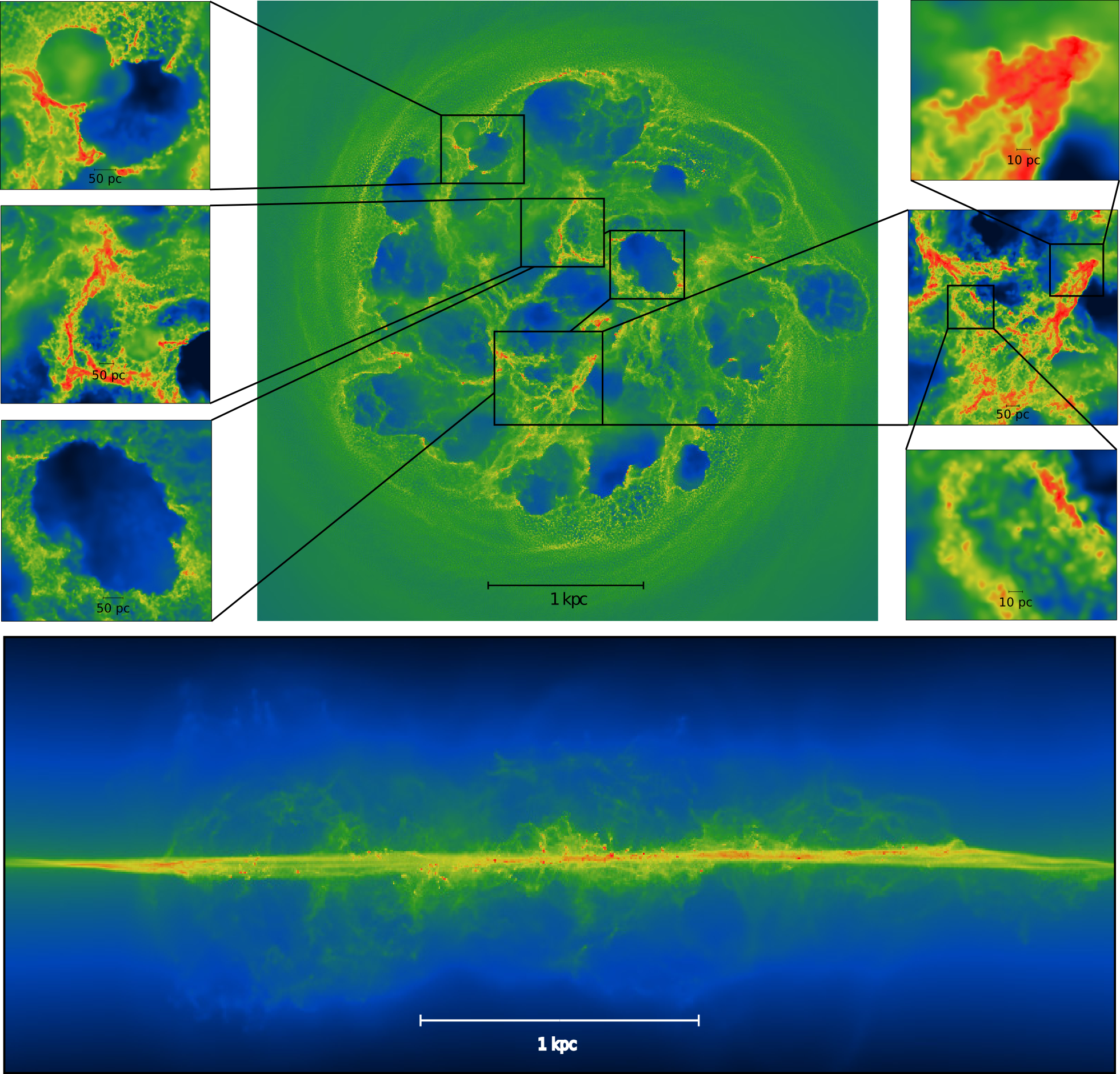}
        \caption{Simulation of an isolated dwarf galaxy with the MFM-solver at a mass resolution of 1 M$_{\odot}$ and a spatial resolution of $0.1$ parsec. The color-coding indicates total gas surface density (covering 10$^{16}$ (blue) to 10$^{22}$ (red) cm-2). We show various zoom-ins on overlapping SN-remnants (top left), filaments (middle panels), an isolated SN-remnant (bottom left), dense cores (top right) and fragmented clouds (bottom right). In the bottom panel we show the edge-on view of the galaxy with filamentary outflows driven by SNe. In simulation the main driver for the resulting complex ISM-structure is the combined effect of the turbulence driving and heating by the feedback of supernovae and the cooling due to the employed non-equilibrium cooling and chemistry routines. The stars in this simulations represent single stars and are sampled with the IMF-sampling approach from \citet{Hu2016}. At the end of their lifetimes the massive stars explode in a core collapse SN-event and distribute 10$^{51}$ erg into the ambient ISM. Every single remnant can be tracked and undergoes a resolved ST-phase in which it generates momentum to drive turbulence and hot mass that is needed to self-consistently provide the pressure support for driving galactic outflows.}
        \label{fig:griffin}
\end{figure*}

One major achievement that can be realised at the target resolution is a full galactic disc simulation that resolves the feedback of massive stars, with the focus on SN-feedback of Type II core collapse supernovae. In Figure \ref{fig:griffin} we show a simulation of an isolated dwarf galaxy that has been simulated with our code and the feedback scheme that we have tested into depth in the scope of this work to investigate the momentum and hot-phase generating properties that can be obtained with a very simple SN-feedback scheme at target resolutions where the the single feedback events' ST-phase is resolved. This kind of simulation shows that SN-feedback can be effective once the ST-phase can be resolved in the majority of the ambient media in which the supernova occurs. With this simple, but resolved feedback scheme it becomes possible to self-consistently study ISM-turbulence and galactic scale outflows in a galaxy scale simulation, without the need for more complicated and less accurate treatments of sub-grid feedback physics, like direct momentum input or a mixed injection scheme that assumes the ST-solution by construction. For our scheme we are able to show that the vast majority of SN-events occurs in environmental densities for which our isolated blast wave studies clearly indicate convergence on momentum and hot phase generation at a mass resolution of 1 M$_{\odot}$. Together, with the results of our presented SN-driven ISM-environments it becomes possible to also achieve hot phase-dominating Volume filling fractions in a galactic scale simulation. Thus far, this has only been achieved in small galactic patches. Finally, we note that with this kind of galaxy-scale simulation we are able to achieve one major step forward in the modelling of ISM-physics in a galactic context. Although, we treat the stars that we form as single stars by randomly drawing them from the IMF, this still represents a sub-grid model for star formation. We then follow the massive stellar population in terms of photo-ionising radiation and photo-electric heating in a Stroemgren-approximation \citep[e.g.][]{Hopkins2011} by mapping the density distribution with a healpix-algorithm and calculating the local column densities with \textsc{treecol} \citep{Clark2012}. Although, we neglect re-emission in this context, we still include an approximation for radiation of the massive stars. Once these massive stars reach the end of their life times they explode as core collapse SNe and every single remnant is able to provide its momentum and hot-phase in the ST-phase removing the sub-grid constraint for supernova-feedback from our galaxy simulations.

\section{Conclusions}
\label{sec:conclusion}
\subsection{Summary}

We carried out three dimensional simulations of isolated SNe that include a non-equilibrium cooling and a  chemical model that tracks the formation (and destruction) of H$_{2}$. The canonical SN-energy of $10^{51}$ erg is coupled by pure thermal injection to one kernel size the centre of a box with uniform density and tested four different mass resolutions of $0.1$ M$_{\odot}$, $1$ M$_{\odot}$, $10$ M$_{\odot}$ and $100$ M$_{\odot}$. The energy is distributed weighted by the kernel. We focus on the convergence of physical properties at the end of the ST-phase to constrain the resolution requirements for a SN-feedback scheme that can be applied in high resolution simulations of galaxy formation and evolution. Specifically, we tested the behaviour of two different, widely used numerical methods for solving Euler's equations. We tested the implications of this feedback scheme with a modified 'pressure-energy' SPH solver and the higher order meshless finite mass (MFM) solver which utilises a second order reconstruction of Euler's equations by solving the Riemann problem on the one dimensional surface between two fluid tracers to obtain the fluid fluxes. The usage of the chemical model allows us to follow the formation and destruction processes of molecular, neutral and ionised gas within isolated SN-remnants as they currently can be modelled in simulations of galaxy formation and evolution. We carried out reference runs with thermal conduction to investigate the effects of the heat flux between SN-bubble and the cold material in the remnant.

\begin{enumerate}

\item \textit{Morphology of the shock:} We can capture three stages in the evolution of a SN-remnant. In the ST-phase we resolve the adiabatic regime defined by the Rankine-Hugoniot jump conditions with an accuracy of up to $10$ per cent that is limited by the behaviour of our underlying numerical scheme. After the ST-phase the remnants enters the PDS-phase in which they can further increase momentum due to the high pressure in the bubble behind the shell. This pushes the shell further outwards until pressure equilibrium with the shell is reached and the shell moves forward with roughly constant momentum and reaches the MCS-phase. Finally, the remnant starts to merge with the ambient ISM. We can capture the shocks morphology reasonably well at our three highest resolution levels. However, at the lowest resolution level the shock remains unresolved.  

\item \textit{Dependence on the environment:} We carried out explosions for six different environmental densities of SN-remnants alongside the equilibrium cooling curve. For all these remnants we computed the end of the ST-phase and find a power law scaling of the termination time of the ST-phase which is within a 30 per cent agreement compared to the work of \citet{Blondin1998}, \citet{Petruk2006} and \citet{Haid2016}. We investigate eight physical quantities as a function of time including radius, momentum, energy distribution, velocity structure, remnant mass, bubble-mass, shell-mass as well as the bubble temperature. At the end of the ST-phase our resolved runs agree very well with the results of studies of the same kind \citep[e.g.][]{Kim2015, Haid2016}, although we note that our default metallicity is $0.1$ Z$_{\odot}$ because we aim for understanding SN-feedback in low metallicity environments. Most quantities can be resolved at a resolution of $10$ M$_{\odot}$ in all relevant density regimes of the ISM. We specifically highlight this in the context of the momentum that has been generated during the ST-phase and the generation of the hot phase (the hot mass and the temperature evolution in the bubble). The feedback can then be resolved as a combination of those two quantities. Momentum is needed to move particles through the ISM and generate turbulence. The hot phase is needed to generate pressure in the ISM which is necessary to launch galactic winds that can impact the CGM of a galaxy. Further, we determine a fit to the SN-momenta that are injected in different regions of the ISM based on our highest resolution remnants which can be adopted for feedback schemes in simulations of galaxy formation and evolution that do not resolve the ST-phase explicitly.

\item \textit{Chemistry of single SN-remnants:} We investigate the chemical composition in the simulated SN-remnants. We note that the remnants in the low density environments contribute to the build up of the hot phase but the remnants become fully neutral after a few cooling times. They do not contribute to the formation of molecular hydrogen because the densities in the swept up mass remain low. In the CNM the mass in the swept up region is fully ionised. After cooling sets in the remnant becomes neutral after a few cooling times again. However at the end of the simulation (t=1 Myr) we find around one per cent of the remnant mass in neutral hydrogen. In the CMM the SN destroys most of the molecular hydrogen and leads to an ionisation of the swept up mass. We find that a small fraction (around $10$ per cent) of the remnants mass remain in the neutral hydrogen phase. After cooling set in most of the material cools quickly and recombination with the free electrons leads to the formation of neutral hydrogen.

\item \textit{Role of thermal conduction:} Thermal conduction leads to a decrease in temperature and to an increase in density in the shell. Thermal energy can be transported from the hot phase to the cold phase while mass from the cold phase is transported to the hot phase. Once cooling becomes relevant the mass flux towards the hot bubble is suppressed because the thermal energy that is transported from the bubble to the shell is instantly radiated away. Thermal conduction makes cooling slightly more efficient and leads to slightly less momentum generation (by $10$ per cent). The quantity that is most affected by thermal conduction is the peak hot mass within the hot bubble which decreases by roughly $40$ per cent compared to the runs without conduction. Due to an increase of the densities within the shell the formation rate of molecular hydrogen increases roughly by a factor of two. Moreover, we find good agreement of our results with the recent work of \citet{El-Badry2019} who investigated the effects of cooling and thermal conduction in simulations of one-dimensional super-bubbles.   

\item \textit{Applications of the presented feedback scheme:} The presented feedback scheme is perfectly suited to be used in future simulations of galaxy formation and evolution that focus on sufficient modelling of the momentum build-up and hot phase generation that can be obtained at a target resolution of a few solar masses to obtain convergence on these quantities over a dynamic range of at least six orders of magnitude. While we extensively tested the properties that can be obtained in isolated unstructured ambient media, we also investigate the case of a feedback event in a highly supersonic, turbulent medium to investigate the momentum and hot-phase build-up in a more realistic environment. Moreover, we investigate the effect of multiple SNe in a closed box model to accurately derive the Volume filling factors for the hot-phase that can be obtained with the feedback scheme. The volume filling hot phase is of importance for the explanation of large scale galactic outflows as we already discussed in section \ref{sec:semi}. We find a dominating hot and warm volume filling fraction for our highest mass resolutions while we generally find that at lower mass resolution the hot volume filling fraction remains low. Thus to self-consistently drive galactic outflows with the feedback that we present it is necessary to have a mass resolution higher than $10$ solar masses.   
\end{enumerate}

\subsection{Model limitations}

Although we find good agreement of our results with other theoretical studies we are still limited by some assumptions we already made in section \ref{sec:methods}. For the isolated remnants we assumed that they occur in an environment with a constant ambient density. Given the highly turbulent structure of the ISM \citep[e.g.][]{Elmegreen2004} this is a rather simplified assumption. Further, we excluded the fact that massive stars normally shape their environment and ionise the surrounding ambient medium due to stellar winds prior to the core collapse event, which renders our assumption of a core collapse event in a fully molecular medium untrue. However, because in this regime we initially destroy all of the molecular hydrogen it does not influence the formation rate after shell formation. Further we neglected that each remnant ejects mass in the explosion, which initially alters the chemical composition of the remnant prior to shell formation. The injection of metals and dust in the explosion would significantly shorten the cooling time and increase the formation rate of the molecules. In a galactic context we can resolve the momentum and the hot phase in the Sedov-Taylor phase, while other properties of the a SN-remnant cannot be captured accurately. For example at the presented mass resolutions it is not possible to reach the predicted analytic value of the density in the adiabatic phase (before cooling dominates) of the remnant which can certainly change the on the fly updated cooling and molecular formation rates.
Further, we do not consider magnetic fields in the shock. This could change the picture again because of the additional pressure component (the same is valid or cosmic rays). 
Moreover, in the case of the runs including thermal conduction it remains unclear to which degree the behaviour is driven by the numerics of the underlying scheme. Here, the crucial point is to resolve the structure of the hot and the cold phase which we do in our two highest resolution runs. If this can be done the effect is physical, if not it remains unclear whether the results have any physical impact. Further, we note that our resolution is far too low to resolve the internal heat fluxes within the bubble or the shell (once it has formed). 

\section*{Acknowledgements}

We thank the referee S.A.E.G. Falle for careful comments on the draft which significantly improved the quality of the paper. UPS thanks Eirini Batziou, Andreas Burkert, Chris Byrohl, Klaus Dolag, Joseph O'Leary, Rhea Silvia Remus, Felix Schulze and Simon White for helpful discussions and insights on the physics of the interstellar medium. UPS thanks Volker Springel and Prateek Sharma for useful discussion on thermal heat conduction. UPS thanks Jeong-Gyu Kim for his comments on the ISM in equilibrium. The authors gratefully acknowledge the computing time granted by the c2pap-cluster in Garching under the project number pr27mi where most of this work has been carried out. UPS and BPM are funded by the Deutsche Forschungsgemeinschaft (DFG, German Research Foundation) with the project number MO 2979/1-1. UPS and TN acknowledge support from the Deutsche Forschungsgemeinschaft (DFG, German Research Foundation) under Germanys Excellence Strategy -
EXC-2094 - 390783311 from the DFG Cluster of Excellence
"ORIGINS". The Flatiron Institute is supported by the Simons Foundation. SWG acknowledges support by the European Research Council via ERC Starting Grant RADFEEDBACK (no. 679852) and by the German Science Foundation via CRC956, Project C5.





\bibliographystyle{mnras}
\bibliography{paper}


\appendix

\section{Different regions of the ISM}
\label{sec:appendixA}

\begin{figure*}
        \includegraphics[scale=0.4]{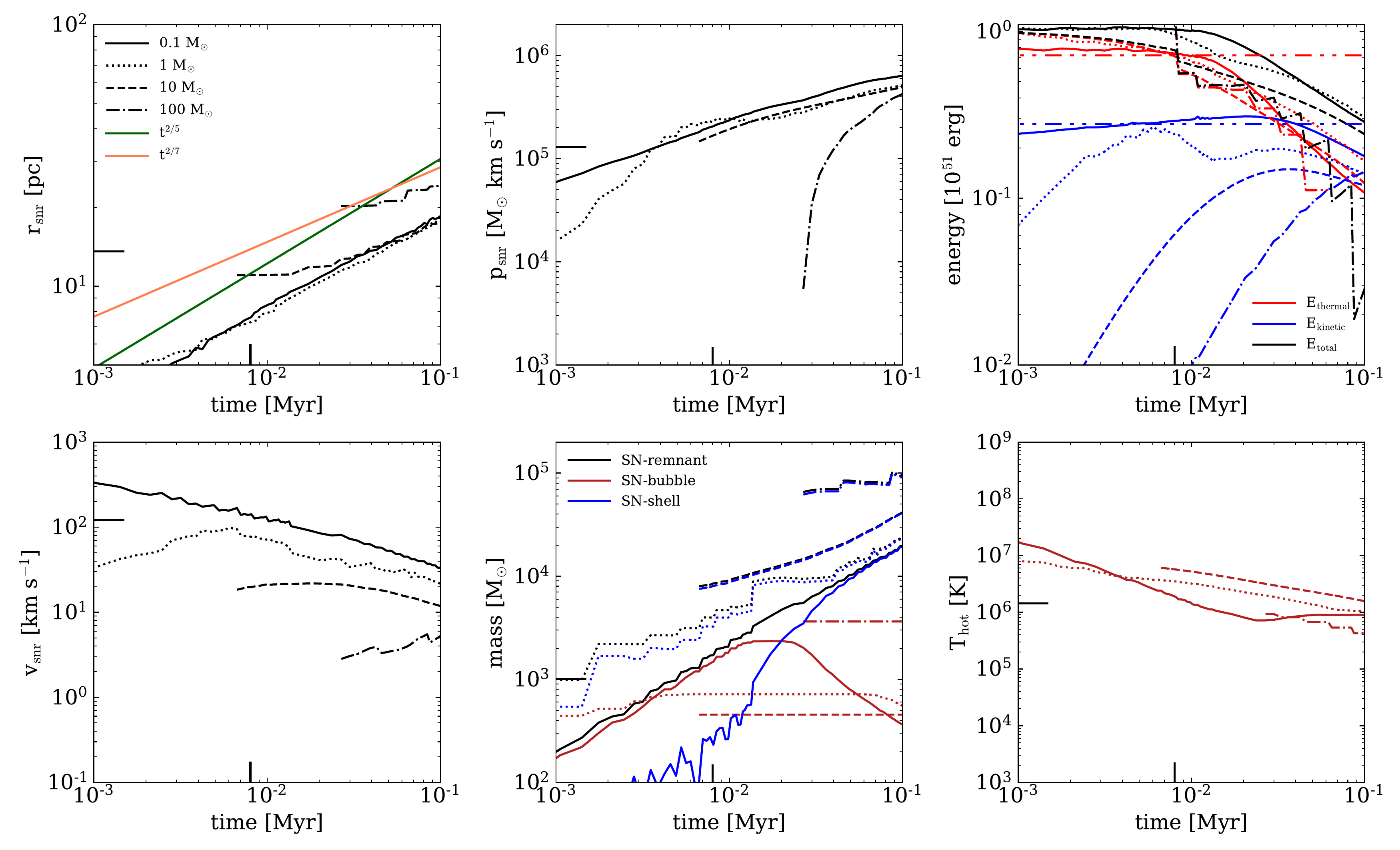}
        \caption{Same as Figure \ref{fig:1cmregion} for an ambient density of $100$ cm$^{-3}$}
        \label{fig:100cmregion}
\end{figure*}

\begin{figure*}
        \includegraphics[scale=0.4]{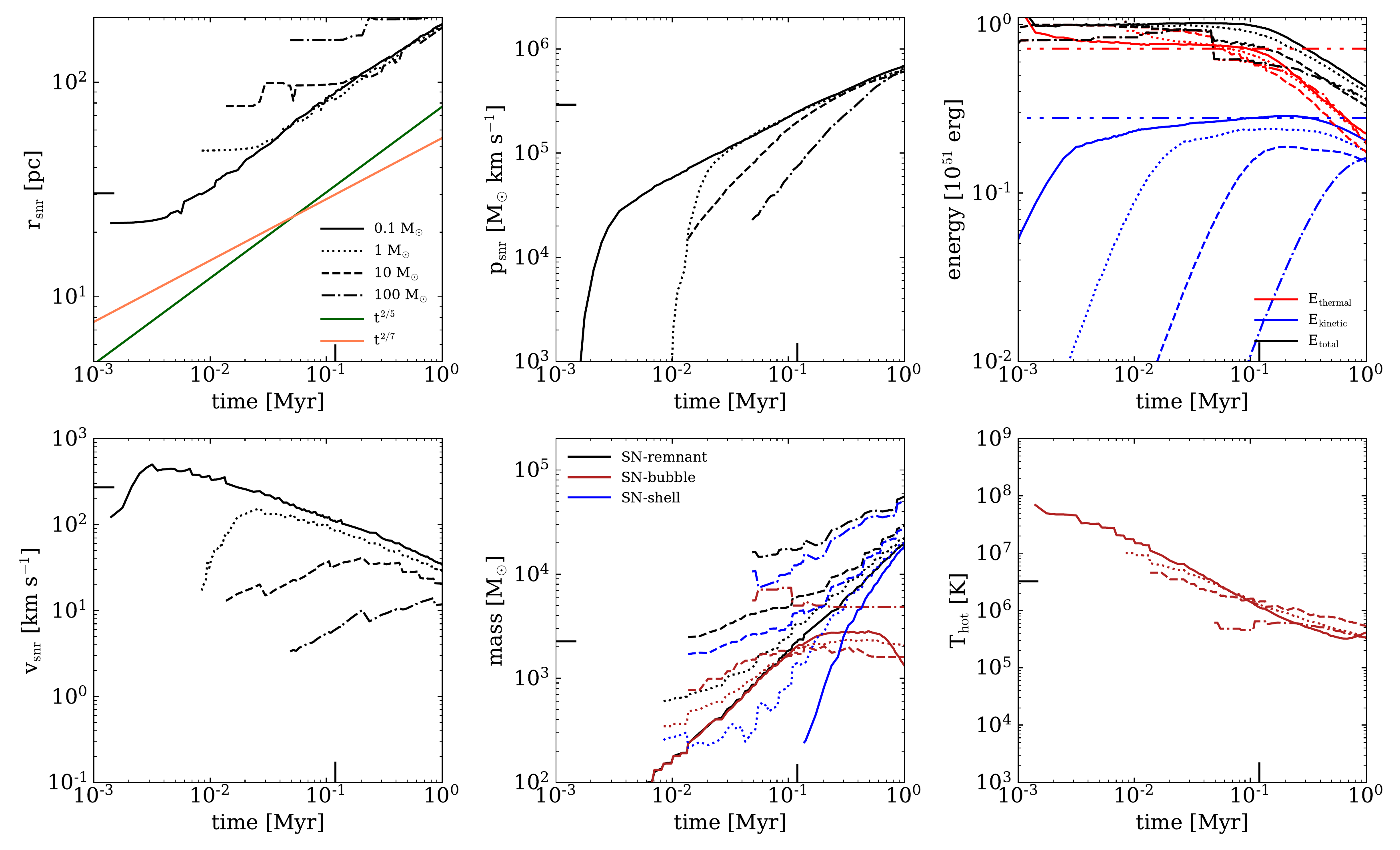}
        \caption{Same as Figure \ref{fig:1cmregion} for an ambient density of $0.1$ cm$^{-3}$}
        \label{fig:01cmregion}
\end{figure*}

In this section we show the results for the runs CMM (Figure \ref{fig:100cmregion}) and WNM (Figure \ref{fig:01cmregion}) with the MFM-solver to investigate the solvers behaviour in capturing the ST-phase as a function of environment. For the higher density environment it is much more difficult to resolve the ST-phase at low mass resolution. While most properties are still within 50 per cent in comparison to \citet{Kim2015} the hot phase remains less and less resolved at low mass resolution. The terminal momentum is still well resolved even at lower resolution. However, as we pointed out in section \ref{sec:resolution_effects} this is a result of the feedback scheme where the thermal energy is injected into the $32$ nearest neighbours which leads to an overestimate of the shell momentum that scales with the particle mass. Therefore, the terminal momentum remains a weak measure to determine whether the ST-phase is resolved or not. Even, if 'correct' terminal momentum is injected and particles move through the volume the feedback remains ineffective as long as the temperatures in shell and bubble are unresolved which then generate the pressure in the ISM. 
For the low density environments all important physical quantities can be properly captured with a mass resolution of 10 M$_{\odot}$. The crucial point in this regime is the long cooling times which are longer than $0.1$ Myr. Because of this long cooling time and the negligible cooling losses it the hot phase in the bubble has time to build up and can be resolved even at lower mass resolutions. As already pointed out above the momentum seems to be converged between all mass resolutions in this environment which is again due to the fact that we sweep up too much mass with our feedback scheme within the ST-phase which counterbalances the poorly resolved velocity structure of the shell which leads to the correct momentum even at lowest mass resolution. However, because in this regime we also revolve the temperature structure of the shell and the bubble the feedback remains resolved because it generates enough pressure within the ISM.     


\bsp	
\label{lastpage}
\end{document}